\documentclass[aps,prd,onecolumn,a4paper,superscriptaddress,nofootinbib]{revtex4-1}
\usepackage{tabularx, amsmath, amssymb, graphicx, multirow, dcolumn, bm, latexsym, soul, nicefrac, subfigure, graphicx, appendix,acronym}
\usepackage[colorlinks, linkcolor=blue, citecolor=blue, urlcolor=blue ]{hyperref}
\usepackage{ulem} 

\begin{document}

\newcommand{\ls}[1]{\textcolor{blue}{LS: #1}}
\newcommand{\yh}[1]{\textcolor{red}{YH: #1}}
\newcommand{\mnras}{Monthly Notices of the Royal Astronomical Society}
\newcommand{\apjl}{the Astrophysical Journal Letters}
\newcommand{\apjs}{the Astrophysical Journal Supplement}
\newcommand{\araa}{Annual Review of Astronomy and Astrophysics}
\newcommand{\aapr}{Astronomy and Astrophysics Reviews}
\newcommand{\pasa}{Publications of the Astronomical Society of Australia}
\newcommand{\aap}{Astronomy and Astrophysics}
\newcommand{\msun}{M_\odot}
\newcommand{\mbh}{M_{\rm BH}}
\newcommand{\mpbh}{M_{\rm PBH}}
\newcommand{\Sa}{\,{\rm m}\,{\rm s}^{-2}\,{\rm Hz}^{-1/2}}
\newcommand{\Sx}{\,{\rm m}\,{\rm Hz}^{-1/2}}
\def\ccr#1{{\color{red}{\bf #1}}}

\title{Science with the TianQin observatory: Preliminary results on stellar-mass binary black holes}

  \author{Shuai Liu }
\email{liush229@mail2.sysu.edu.cn}
\affiliation{TianQin Research Center for Gravitational Physics and School of Physics and Astronomy, Sun Yat-sen University (Zhuhai Campus), Zhuhai 519082, People's Republic of China}

  \author{Yi-Ming Hu }
\email{huyiming@sysu.edu.cn}
\affiliation{TianQin Research Center for Gravitational Physics and School of Physics and Astronomy, Sun Yat-sen University (Zhuhai Campus), Zhuhai 519082, People's Republic of China}

  \author{Jian-dong Zhang }
\email{zhangjd9@mail.sysu.edu.cn}
\affiliation{TianQin Research Center for Gravitational Physics and School of Physics and Astronomy, Sun Yat-sen University (Zhuhai Campus), Zhuhai 519082, People's Republic of China}

  \author{Jianwei Mei }
\email{meijw@sysu.edu.cn}
\affiliation{TianQin Research Center for Gravitational Physics and School of Physics and Astronomy, Sun Yat-sen University (Zhuhai Campus), Zhuhai 519082, Peoples's Republic of China}
\affiliation{MOE Key Laboratory of Fundamental Physical Quantities Measurements, Hubei Key Laboratory of Gravitation and Quantum Physics, PGMF and School of Physics, Huazhong University of Science and Technology, Wuhan 430074, Peoples's Republic of China}
\date{\today}

\begin{abstract}

We study the prospect of using TianQin to detect stellar-mass binary black holes (SBBHs).
We estimate the expected detection number as well as the precision of parameter estimation on SBBH inspirals, using five different population models.
We note TianQin can possibly detect a few SBBH inspirals with signal to noise ratios greater than 12; lowering the threshold and combining multiple detectors can both boost the detection number.
The source parameters can be recovered with good precision for most events above the detection threshold.
  For example, the precision of the merger time most likely occurs near 1s, making it possible to guide the detection of the ground-based detectors, the precision of the eccentricity $e_0$ most likely occurs near $10^{-4}$, making it possible to distinguish the formation channels, and the precision of the mass parameter is better than $10^{-6}$ in general and most likely occurs near $10^{-7}$.
  We note, in particular, that for a typical merger event, the error volume is likely to be small enough to contain only the host galaxy, which could greatly help in the study of gravitational wave cosmology and relevant studies through the multimessenger observation.
\end{abstract}

\maketitle

\acrodef{GW}{gravitational wave}
\acrodef{LIGO}{Laser Interferometer Gravitational-Wave Observatory}
\acrodef{aLIGO}{advanced LIGO}
\acrodef{AdV}{advanced Virgo}
\acrodef{O1}{the first observational run}
\acrodef{O2}{the second observational run}
\acrodef{O3}{the third observational run}
\acrodef{BBH}{binary black hole}
\acrodef{BNS}{binary neutron star}
\acrodef{BH}{black hole}
\acrodef{XRB}{x-ray binary}
\acrodef{EM}{electromagnetic}
\acrodef{SBH}{stellar-mass black hole}
\acrodef{PBH}{primordial black hole}
\acrodef{SBBH}{stellar-mass binary black hole}
\acrodef{SNR}{signal-to-noise ratio}
\acrodef{LISA}{Laser Interferometer Space Antenna}
\acrodef{LVC}{LIGO Scientific Collaboration and Virgo Collaboration}
\acrodef{IMF}{initial mass function}
\acrodef{PN}{post-Newtonian}
\acrodef{SSB}{solar system barycenter}
\acrodef{SPA}{stationary phase approximation}
\acrodef{PSD}{power spectral density}
\acrodef{FIM}{Fisher information matrix}
\acrodef{PPISN}{pulsational pair-instability supernova}

\section{Introduction}
The gravitational collapse of massive stars can produce \acp{SBH} with a mass range from a few to one hundred solar masses \cite{Burrows:1988ba, OConnor:2010moj, Colpi:2016fup}. Other mechanism can also produce black holes with similar masses, for example, \acp{PBH} may result from the discontinuity or unevenness of matter distribution in the very early universe, and the \acp{PBH} channel for the formation of \acp{SBH} cannot yet be fully excluded by observations \cite{Bird:2016dcv, Carr:2016drx, Sasaki:2016jop, Ali-Haimoud:2017rtz, Inomata:2016rbd, Ando:2017veq, Chen:2018czv, Sasaki:2018dmp}.

Before the first \ac{GW} detection of \acp{SBBH} coalescence by the \ac{LVC} \cite{Abbott:2016blz}, \acp{SBH} can only be observed by effects induced on their companions as well as through their accretion process with the \ac{EM} observations.
More specifically, with the \ac{EM} channel, our understanding about \acp{SBH} came mostly from the investigations of \ac{XRB}, which consists of a stellar object and an accreting compact object.
At present, 22 \acp{XRB} containing \acp{SBH} have been observed \cite{TheLIGOScientific:2016htt}, with most of the \acp{SBH} in these systems being lighter than $20M_{\odot}$.
\acp{SBH} with higher masses have been claimed for detection, but mainly they remain under debate \cite{2019Natur.575..618L}.
Observations of \acp{SBH} in \ac{EM} channels indicated that there might be a gap between the most massive neutron stars \cite{Freire:2007jd, Ozel:2016oaf, Margalit:2017dij} and the lightest \acp{BH} with mass at about $5\msun$ \cite{Ozel:2010su, Farr:2010tu, Kreidberg:2012ud}. 
It was proposed that the existence of this gap could be constrainted by \ac{GW} observation \cite{Littenberg:2015tpa, Mandel:2015spa, Mandel:2016prl, Kovetz:2016kpi}( {\it cf.} in \ac{O3} \ac{LVC} reported possible observations of compact objects in this mass gap \cite{S190924h,S190930s,S191216ap}). 
Before the \ac{GW} detection, there were large uncertainties on the merger rates of \acp{SBBH}, ranging from $0.1$ to $\sim 300{\rm Gpc^{-3}yr^{-1}}$ \cite{Abadie:2010cf, Downing:2009ag, Downing:2010hq, Mennekens:2013dja, Dominik:2014yma, Rodriguez:2015oxa, Mandel:2015qlu}.

The understanding of \acp{SBH} was revolutionized on September, 14$^{\rm th}$, 2015, when the first \ac{GW} signal from the merging stellar-mass \ac{BBH}, later named as GW150914, was detected by the \ac{aLIGO} detectors\cite{Abbott:2016blz,2016PhRvL.116f1102A,2016PhRvL.116m1103A,2016PhRvD..93l2003A,2016PhRvL.116x1102A,2016PhRvL.116v1101A,2016ApJ...833L...1A,2016PhRvD..93l2004A,2016CQGra..33m4001A,2017PhRvD..95f2003A,2016ApJ...818L..22A,2016PhRvL.116m1102A}.
A new window of \ac{GW} has been opened to observe the Universe since.
Further analysis reveals that the component masses of GW150914 are $35.6^{+4.8}_{-3.0}\msun$ and $30.6^{+3.0}_{-4.4}\msun$ respectively, with a redshift of $z=0.09^{+0.03}_{-0.03}$ \cite{LIGOScientific:2018mvr}.
The event, followed by nine other announced detections of \acp{SBBH} mergers and one announced detection of \ac{BNS} merger by \ac{aLIGO} and Virgo in the \ac{O1} and \ac{O2}, marked the beginning of \ac{GW} astronomy \cite{2016PhRvX...6d1015A,2016PhRvL.116x1103A,2016PhRvX...6d1015A,2017PhRvL.118v1101A,2017ApJ...851L..35A,2017PhRvL.119n1101A,2017PhRvL.119p1101A,2017ApJ...848L..12A,2017ApJ...848L..13A,2017Natur.551...85A}.
In the \ac{O3}, \ac{LVC} published the detections of GW190412 and GW190425 \cite{Abbott:2020uma, LIGOScientific:2020stg}, and a list of compact binary coalescence alerts were released to public promptly. A more in-depth analysis is yet to be published.
In the future, more detectors, like KAGRA \cite{Aso:2013eba}, are also aiming to join the collaboration and contribution.

The component masses of many \acp{SBH} observed by \ac{LVC} are greater than $20M_{\odot}$, which is systematically larger than those in \acp{XRB} \cite{LIGOScientific:2018mvr, LIGOScientific:2018jsj}.
Meanwhile, the direct observation of BBH mergers greatly improve our understanding on their event rates.
The observation of GW150914 alone constraints the rate to be  $2-400{\rm Gpc^{-3}yr^{-1}}$ \cite{Abbott:2016nhf}, while combining all events detected in \ac{O1} and \ac{O2} further shrinks the range to $25-109{\rm Gpc^{-3}yr^{-1}}$ \cite{LIGOScientific:2018jsj}.
All observed individual black holes masses are consistent with the theoretical upper limit of $\sim 50\msun$ induced by the \ac{PPISN} and pair-instability supernova \cite{Heger:2001cd, Belczynski:2016jno, Woosley:2016hmi, Spera:2017fyx, Marchant:2018kun}.

The observation of \acp{SBBH} mergers poses two questions: how do \acp{SBH} form and how do they bind into binaries.
\acp{SBH} may originate from three scenarios:
(i) The collapse of massive stars, which depends strongly on the star's metallicity, stellar rotation, and the microphysics of stellar evolution, which metallicity has the greatest impact.
Lower metallicities lead to weaker stellar winds and can result in the formation of more massive \acp{SBH} \cite{Belczynski:2009xy, Mapelli:2012vf, Spera:2015vkd}.
(ii) \acp{SBH} from \acp{PBH} \cite{Bird:2016dcv, Clesse:2016vqa, Sasaki:2016jop, Kashlinsky:2016sdv, Sasaki:2018dmp}.
\acp{PBH} can be formed through several mechanisms, the most popular being the gravitational collapse of overdensity regions \cite{Niemeyer:1997mt, Shibata:1999zs, Musco:2004ak, Polnarev:2006aa, Musco:2008hv, Nakama:2013ica, Harada:2013epa}.
As the mass spectrum for this process can be quite wide, there will be no difficulty for massive \acp{SBH} formed through this channel \cite{Garcia-Bellido:2017mdw, Sasaki:2018dmp}. 
(iii) \acp{SBH} could also be a product of former \ac{SBBH} mergers \cite{OLeary:2016ayz, Fishbach:2017zga, Gerosa:2017kvu, Rodriguez:2017pec,Veske:2020zch} . 

On the other hand, the binding of \acp{SBBH} can be largely categorized into two channels, where great progress has been made since the first \ac{GW} detection, which is also reflected from the update of esitmated rates.
(i) Coevolution of massive star binaries (e.g., \citep[][]{Vanbeveren:2008sj, Belczynski:2010tb, TheLIGOScientific:2016htt, Kruckow:2018slo,Giacobbo:2018etu}), with the corresponding merger rates ranging from $6$ to $240{\rm Gpc^{-3}yr^{-1}}$ (e.g., \citep[][]{Mapelli:2018wys, Buisson:2020hoq}).
In this scenario, \acp{SBBH} will inherit the orbits and spins of their stellar progenitors; frictions within common envelope and other late stellar evolution process will shrink the eccentricities and align component of \ac{SBH} spin to the orbital angular momentum.
(ii) Dynamical process in dense stellar environments (e.g., \citep[][]{PortegiesZwart:1999nm, Gultekin:2004pm, Gultekin:2005fd, TheLIGOScientific:2016htt, Chatterjee:2016thb, Zevin:2018kzq, Tagawa:2019osr, Samsing:2019dtb}), with a meger rate estimation within the range of $5-70{\rm Gpc^{-3}yr^{-1}}$ (e.g., \citep[][]{Rodriguez:2016kxx, Rodriguez:2016avt, Park:2017zgj, Kumamoto:2020wqr}).
The dynamical nature of their origin would implicate a relatively large orbital eccentricities as well as an isotropic distribution of the spins for the component \acp{SBH} \cite{Samsing:2013kua, Antonini:2015zsa, LIGOScientific:2018jsj, Samsing:2018isx, Kremer:2018cir}. 
Moreover, if \acp{SBBH} are composed of \acp{PBH}, then they are also expected to be formed through the dynamical encounter process \cite{Sasaki:2016jop, Bird:2016dcv}; the merger rates depend on the fraction of \acp{PBH} in dark matter and mass function (e.g., \citep[][]{Raidal:2017mfl, Chen:2018czv, Raidal:2018bbj}). These \acp{SBBH} are also expected to have large orbital eccentricities \cite{Cholis:2016kqi}.

While \acp{SBBH} merge at high frequencies where ground-based \ac{GW} detectors are most sensitive, \ac{GW} signals from their early inspiral could be observed by space-borne \ac{GW} detectors with sensitive frequencies at millihertz range.
By adopting a nominal detection threshold of \ac{SNR} equal to 8, several studies claim that eLISA/LISA could individually resolve up to thousands of \acp{SBBH} \cite{Sesana:2016ljz, Sesana:2017vsj, Seto:2016wom, Kyutoku:2016ppx}, and the detection capability of Pre-DECIGO (recently renamed as B-DECIGO) has also been investigated \cite{Nakamura:2016hna,Isoyama:2018rjb}. 
New proposals for future generation space-borne \ac{GW} detectors have also been proposed to better observe the \acp{SBBH} \cite{Sedda:2019uro, Kuns:2019upi}. 

With the expected \acp{SBBH} detections from space-borne \ac{GW} detectors, a number of studies have explored their potential to distinguish the formation scenarios of \acp{SBBH}, by measuring the orbital eccentricities \cite{Nishizawa:2016jji, Nishizawa:2016eza, Breivik:2016ddj},
by using imprint of center of mass acceleration of SBBHs on the \ac{GW} signals \cite{Inayoshi:2017hgw, Randall:2018lnh},
and by counting the detection rates \cite{Gerosa:2019dbe, Randall:2019znp}.
Furthermore, if the host galaxy of the \acp{SBBH} can be successfully identified, such systems could also provide a powerful laboratory for cosmology and fundamental physics.
It is argued that \acp{SBBH} detections in millihertz band could be used to study cosmology as standard sirens \cite{DelPozzo:2017kme}.
Reference \cite{Kyutoku:2016zxn} proposes that LISA can use \acp{SBBH} detections to measure the Hubble parameter.
Space-borne detectors could also constrain certain parameters of modified gravity theories with great precisions \cite{Barausse:2016eii, Chamberlain:2017fjl}.

Multiband \ac{GW} astronomy is an important aspect of the science with \acp{SBBH} \cite{Sesana:2016ljz,Seto:2016wom}.
The joint observation of space-borne and ground-based GW detectors could increase the scientific payoff, such as improving the constraint on the source parameters of \acp{SBBH} or on the consistency tests of general relativity \cite{Vitale:2016rfr,Tso:2018pdv} and lowering the detection SNR threshold for space-based detectors by using information from ground-based detectors \cite{Wong:2018uwb}.

The space-based \ac{GW} observatory TianQin is expected to start operation around 2035  \cite{Luo:2015ght}.
In this paper, we focus our attention on the detection ability as well as the precision of parameter estimation of TianQin on \acp{SBBH}.
A collection of five different \acp{SBBH} mass distributions, with corresponding rates inferred from \ac{GW} observations, is adopted.
Based on the detection number as well as parameter estimation calculations, we further investigate the capability of TianQin to provide an early warning, and to explore its potential to multiband \ac{GW} observations and multimessenger studies.
We also discuss the potential of TianQin on astrophysics and fundamental physics with \acp{SBBH}, such as discriminating the formation channels of \acp{SBBH}, etc.

The paper is organised as follows.
In Sec. \ref{sec:distribution}, we introduce the \ac{SBBH} mass distribution models that we need in the study.
In Sec. \ref{sec:method}, we describe the waveform and the statistical method employed.
In Sec. \ref{sec:result}, we present the main results of this work.
In Sec. \ref{sec:summary}, we conclude with a short summary.
Throughout the paper, we use the geometrical units ($G=c=1$) unless otherwise stated.

\section{Mass distribution models}\label{sec:distribution}

\begin{figure}[h]
\centering
\subfigure{
\includegraphics[width=0.48\textwidth]{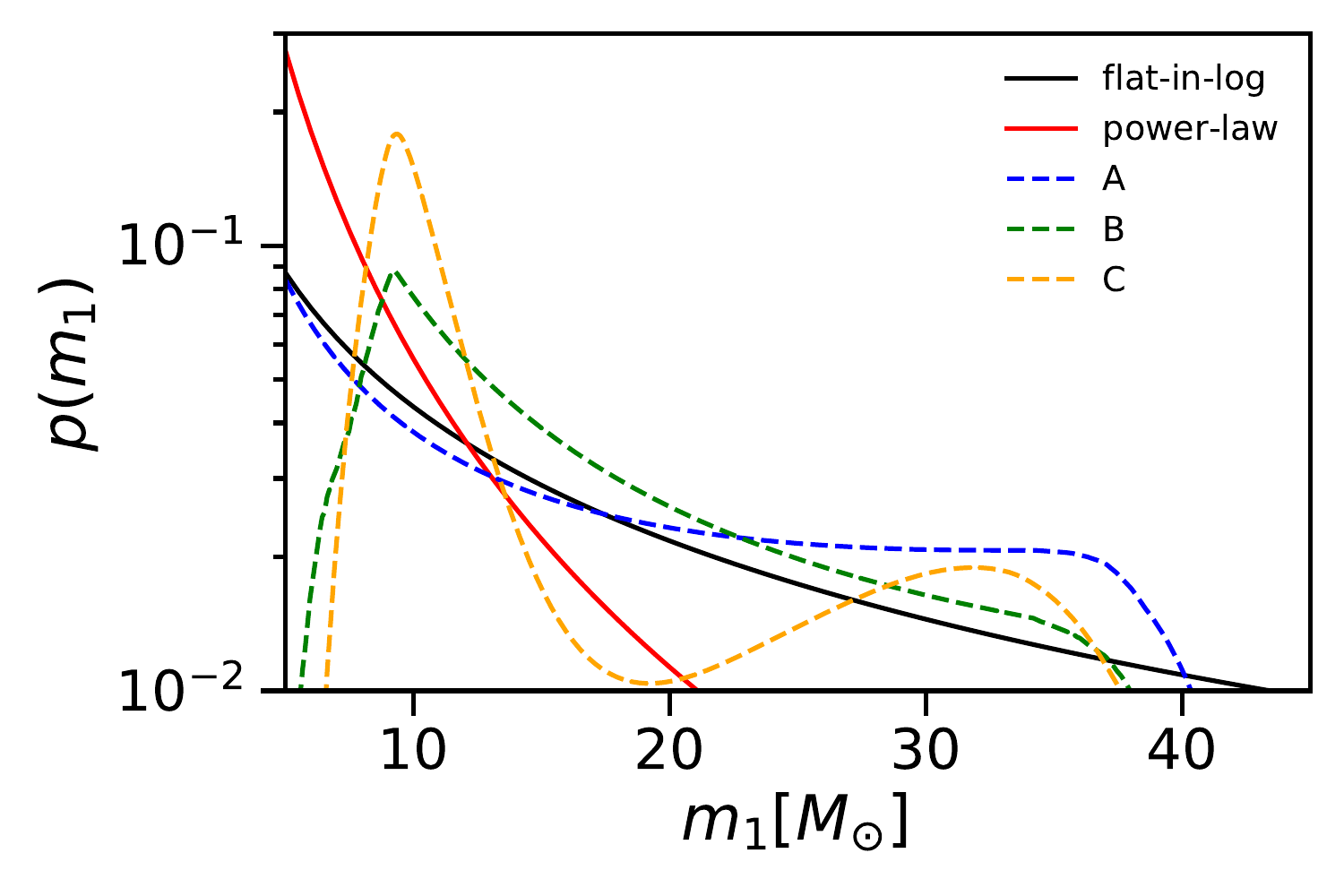}}
\subfigure{
\includegraphics[width=0.48\textwidth]{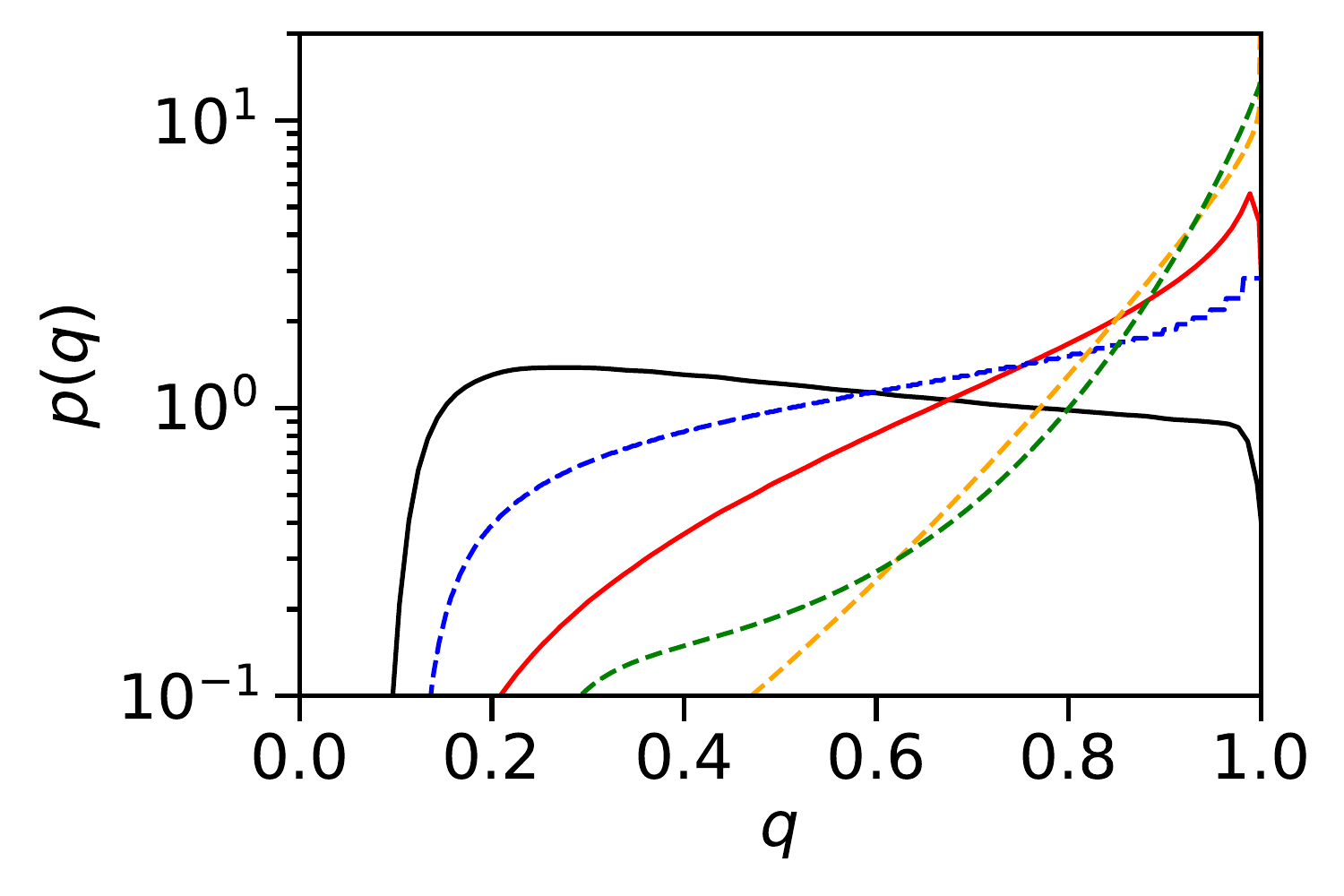}}
\caption{Distribution of the primary mass $m_{1}$ (left) and the mass ratio $q$ (right) for five mass models.
  The black and red solid lines represent models flat-in-log and power law; blue, green, and orange dashed lines denote models A, B, and C, respectively.}
\label{fig:MassDistri}
\end{figure}

Prior to the \ac{GW} detections of \ac{SBBH} mergers, their mass distribution was derived by studying the evolution of massive stars.
Observational evidence indicates that the \ac{IMF} for progenitors is well approximated by a single power law \cite{Salpeter:1955it}; the \emph{power-law} model thus adopts the extreme assumptions that the mass distribution of \acp{SBH} follows closely with the \ac{IMF}, while the other extreme model assumes a \emph{flat-in-log} distribution.
These two extreme models were adopted for relevant calculations \cite{Dominik:2012kk, Fryer:1999ht, Fryer:2011cx, Spera:2015vkd}. During the calculation of merger rate, the selection bias makes the flat-in-log model a pessimistic prediction while the power law an optimistic model in terms of the \acp{SBBH} merger rate \cite{Dominik:2012kk, Fryer:1999ht, Fryer:2011cx, Spera:2015vkd}.
As \ac{GW} observation results accumulate, several phenomenological mass distribution models of \acp{SBBH} are constructed and calibrated: models \emph{A}, \emph{B}, and \emph{C} \cite{LIGOScientific:2018jsj}, taking into consideration the SBH mass gaps \cite{Fishbach:2017zga} and an excess of SBHs with masses near $40\msun$ caused by \ac{PPISN} \cite{Talbot:2018cva}. 
More details of the five models can be found in Appendix \ref{sec:app1}.
For the five models, the distributions of primary mass $m_{1}$ and the mass ratio $q$ are shown in Fig. \ref{fig:MassDistri}.
The obvious outlier is the flat-in-log model, showing a much steeper tail in heavier end.

Previously, the merger rate of \acp{SBBH} $\mathcal{R}$ was mostly obtained through population synthesis, and it span 3 orders of magnitude \cite{Abadie:2010cf}.
With the \ac{GW} detection of \acp{SBBH} by \ac{LVC}, by correcting the selection bias introduced by the assumed underlying mass distribution models, one can derive the corresponding merger rates.
Currently the uncertainty of merger rate has been greatly reduced \cite{Abbott:2016nhf, LIGOScientific:2018mvr, LIGOScientific:2018jsj}.
The merger rate distributions in comoving volume of the five mass distribution models are listed in Table \ref{tab:MergRate} in Appendix \ref{sec:app1} and plotted in Fig. \ref{fig:MergerRate}, respectively.
\footnote{Notice that for simplicity, we do not consider the model evolution against redshift.}
All distributions roughly follow the log-normal distributions.
We note that since the flat-in-log model assumes a much shallower tail than the remaining four models, combined with the selection bias of ground-based \ac{GW} detectors toward higher mass \ac{SBBH} mergers, the same amount of detections would translate into lower overall rates.

\begin{figure}[h]
\centering
\includegraphics[width=0.5\textwidth]{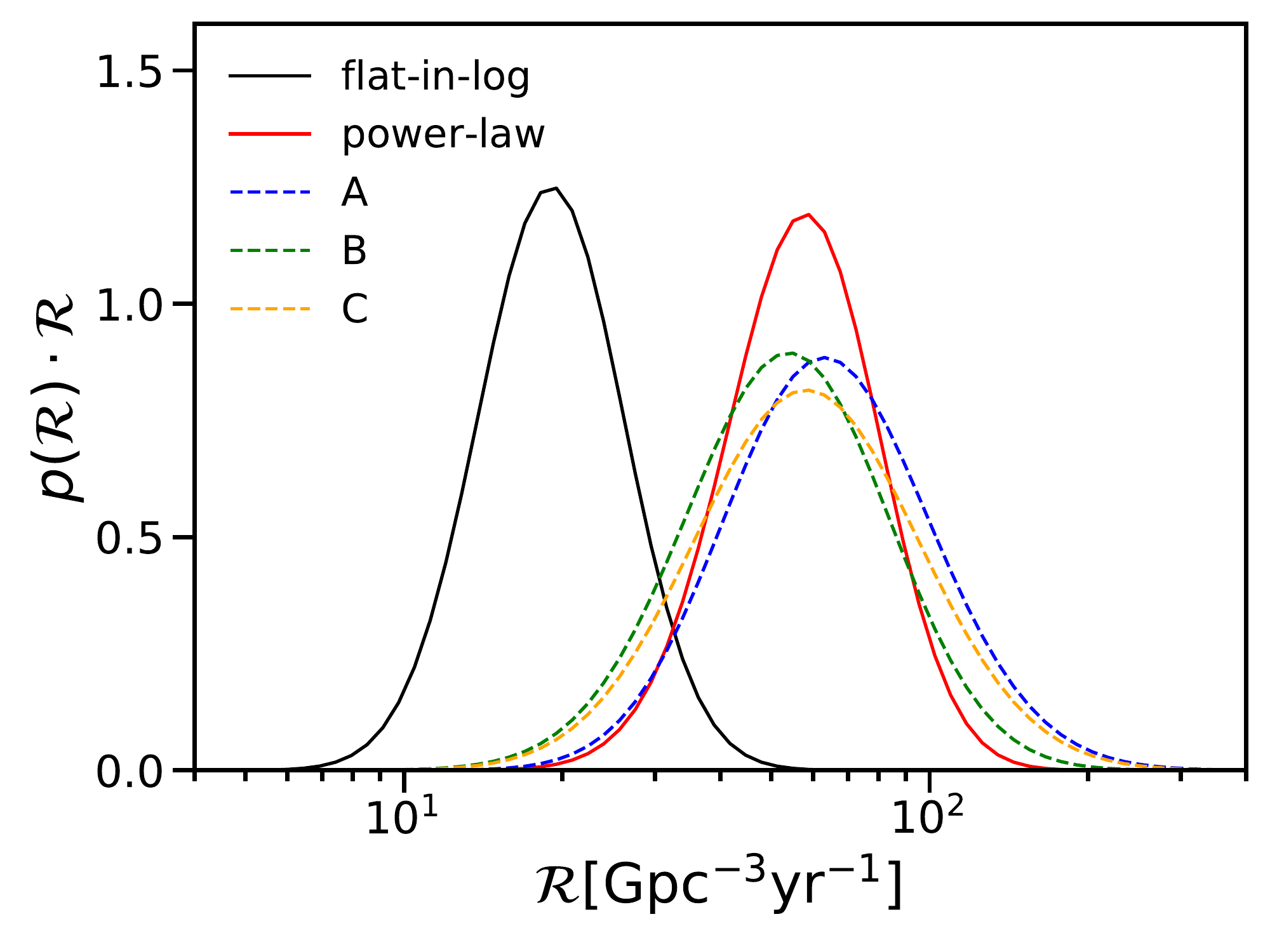}
  \caption{The distributions of merger rates $\mathcal{R}$ of the five mass models.
  The black and red solid lines represent models flat-in-log and power law; blue, green, and orange dashed lines denote models A, B, and C, respectively.}
\label{fig:MergerRate}
\end{figure}

\section{Method}\label{sec:method}
\subsection{Waveform and response}
Since in the mHz range where the space-borne \ac{GW} detectors are most sensitive to, \acp{SBBH} locate well within the inspiral stage; thus, \ac{PN} waveform is sufficient to precisely describe the waveform \cite{Mangiagli:2018kpu}.
For a circular orbit \ac{SBBH} system consisting of two SBHs with masses $m_1$ and $m_2$, the emitted inspiral \ac{GW} in the detector frame can be described as \cite{Colpi:2016fup}
\begin{subequations}\label{eq:h_t}
\begin{align}
  h_{+}(t)&=\frac{2\mathcal{M}^{5/3}(\pi f)^{2/3}(1+\cos^{2}\iota)}{d}\cos\left(\int {\rm d}t\, 2\pi f\right), \\
  h_{\times}(t)&=-\frac{4\mathcal{M}^{5/3}(\pi f)^{2/3}\cos\iota }{d}\sin\left(\int {\rm d}t\, 2\pi f\right),
\end{align}
\end{subequations}
where $f$ is the frequency of GW, $d$ is the distance between the detector and the \ac{SBBH}, $\mathcal{M}=\eta^{3/5}M$ ($\eta=m_{1}m_{2}/M^{2}$ being the symmetric mass ratio and $M=m_{1}+m_{2}$ being the total mass) is the chirp mass, and $\iota=\arccos(\hat{\mathbf{n}}\cdot \hat{\mathbf{L}})$ is the inclination angle of the orbit to the line of sight.
Notice that the redshift $z$ will modify the frequency/time by a factor of $1+z$.
In practice, we replace distance $d$ with the luminosity distance $D_{L}$.
Also, the chirp mass and total mass will be converted to the redshifted quantities in the observer frame \cite{Colpi:2016fup},
\begin{equation}
  \{\mathcal{M}, M\} \rightarrow \{\mathcal{M}(1+z), M(1+z)\}.
\end{equation}
Throughout the paper, relevant mass parameters are referred to the redshifted quantities.

The TianQin satellites are designed to follow a geocentric orbit, with arm length of about $L=\sqrt{3}\times 10^5$ km, performing laser interferometry to detect \ac{GW} signals.
The orientation of the orbital plane is fixed in space.
To cope with the converse effect of the sunlight, the initial design of TianQin opts for a conservative strategy by observing only for every other three months.
The three arms of the TianQin detector can be combined into two independent Michelson interferometers in the low frequency region ($f<f_{\ast}\equiv 1/(2\pi L)\approx0.28{\rm Hz}$), which is generally valid for most of the \acp{SBBH} \cite{hu64tianqin,10.1093/nsr/nwx115}.

For each Michelson interferometer, the detected signal $h(t)$ can be expressed as \cite{Klein:2015hvg}
\begin{align}
  h_{\alpha}(t)&=\frac{\sqrt{3}}{2}\left[ F^{+}_{\alpha}(t)h_{+}(t-t_{D})+F^{\times}_{\alpha}(t)h_{\times}(t-t_{D}) \right], \qquad \alpha = 1, 2 \\
  t_{D}&=R\sin\bar{\theta}_{S}\cos[\bar{\Phi}(t)-\bar{\phi}_{S}]\,,
\end{align}
where $t_{D}$ is the delay time between the interferometer and the \ac{SSB}, $R = 1$ AU and $\bar{\Phi}(t)=\bar{\phi}_{0}+2\pi t/T$, $T=1$~year is Earth's orbital period around the Sun, and $\bar{\phi}_{0}$ is the initial location of TianQin at time $t=0$.
Note the barred variables are quantities in the frame fixed on \ac{SSB} and the unbarred variables are quantities in the detector frame.
In the low frequency region, the antenna pattern functions $F_{\alpha}^{+,\times}(t)$ are \cite{Cutler:1997ta}
\begin{subequations}\label{eq:h_t_response}
\begin{align}
  F_{1}^{+}(t)&=\frac{1}{2}(1+\cos^{2}\theta_{S})\cos2\phi_{S}\cos2\psi_{S}-\cos\theta_{S}\sin2\phi_{S}\sin2\psi_{S}, \\
  F_{1}^{\times}(t)&=\frac{1}{2}(1+\cos^{2}\theta_{S})\cos2\phi_{S}\sin2\psi_{S}+\cos\theta_{S}\sin2\phi_{S}\cos2\psi_{S},\\
  F_{2}^{+}(t)&=F_{1}^{+}(\theta_{S},\phi_{S}-\frac{\pi}{4},\psi_{S}),\\
  F_{2}^{\times}(t)&=F_{1}^{\times}(\theta_{S},\phi_{S}-\frac{\pi}{4},\psi_{S}),
\end{align}
\end{subequations}
where $\theta_{S}$ and $\phi_{S}$ are the altitude and azimuth angle, respectively, of the source.
The polarization angle $\psi_{S}$ is defined as
\begin{equation}
  \tan\psi_{S}=\frac{\mathbf{\hat{L}}\cdot\mathbf{\hat{z}}-(\mathbf{\hat{L}}\cdot\mathbf{\hat{n}})(\mathbf{\hat{z}}\cdot{\mathbf{\hat{n}}})}{\mathbf{\hat{n}}\cdot(\mathbf{\hat{L}}\times\mathbf{\hat{z}})},
\end{equation}
where $\mathbf{\hat{z}}$ is the unit normal vector of the orbital plane of TianQin, $\mathbf{\hat{n}}$ is the unit vector to the source, and $\mathbf{\hat{L}}$ is the unit vector of the angular momentum of the source.
Since we ignore the impact of \ac{BH} spins, the polarization angle $\psi_{S}$ is fixed.

Away from the low frequency region, i.e. for $f>f_{\ast}$, the antenna pattern functions are frequency dependent and they are also complicated to calculate.
In this study, we adopt a common simplification by absorbing such frequency dependence into the detector noise and use (\ref{eq:h_t_response}) for the whole frequency range targeted by TianQin.

As we will see in Sec. \ref{sec:SNR}, it is convenient to perform calculation in the frequency domain.
We express the frequency domain signal $\widetilde{h}_{\alpha}(f)$ as the Fourier transform of the time domain signal\footnote{We note that \ac{SPA} could also be used to derive the \ac{GW} strain after response.
However, \ac{SPA} requires an analytical expression for the waveform, which is valid for \ac{PN} approximations, but not valid for more general cases.
Therefore, we adopt this convolution method, in which we test the validity through numerical comparison between Eq. (\ref{eq:h_f}) and discrete Fourier transform of Eq. (\ref{eq:h_t_response}).}
\begin{equation}\label{eq:h_f}
  \widetilde{h}_{\alpha}(f)=\frac{\sqrt{3}}{2}\left\{\mathcal{F}[h_{+}(t-t_{D})F_{\alpha}^{+}(t)]+\mathcal{F}[h_{\times}(t-t_{D})F_{\alpha}^{\times}(t)]\right\},
\end{equation}
with
\begin{subequations}\label{eq:FT_response}
\begin{align}
  \mathcal{F}[h_{+}(t-t_{D})F^{+}_{1}(t)]&=\frac{1}{4}(1+\cos^{2}\theta_{S})\left[e^{2i\zeta_{1}(f-2f_{0})}\widetilde{h}_{+}(f-2f_{0})+e^{-2i\zeta_{2}(f+2f_{0})}\widetilde{h}_{+}(f+2f_{0})\right]\cos2\psi_{S} \nonumber \\
  &-\frac{i}{2}\cos\theta_{S}\left[-e^{2i\zeta_{1}(f-2f_{0})}\widetilde{h}_{+}(f-2f_{0})+e^{-2i\zeta_{2}(f+2f_{0})}\widetilde{h}_{+}(f+2f_{0})\right]\sin2\psi_{S},\label{eq:8a} \\
  \mathcal{F}[h_{\times}(t-t_{D})F^{\times}_{1}(t)]&=\frac{1}{4}(1+\cos^{2}\theta_{S})\left[e^{2i\zeta_{1}(f-2f_{0})}\widetilde{h}_{\times}(f-2f_{0})+e^{-2i\zeta_{2}(f+2f_{0})}\widetilde{h}_{\times}(f+2f_{0})\right]\sin2\psi_{S} \nonumber\\
  &+\frac{i}{2}\cos\theta_{S}\left[-e^{2i\zeta_{1}(f-2f_{0})}\widetilde{h}_{\times}(f-2f_{0})+e^{-2i\zeta_{2}(f+2f_{0})}\widetilde{h}_{\times}(f+2f_{0})\right]\cos2\psi_{S}, \label{eq:8b}\\
  \mathcal{F}[h_{+}(t-t_{D})F^{+}_{2}(t)]&=\mathcal{F}[h_{+}(t-t_{D})F^{+}_{1}(\phi_{S0}-\frac{\pi}{4})], \\
  \mathcal{F}[h_{\times}(t-t_{D})F^{\times}_{2}(t)]&=\mathcal{F}[h_{\times}(t-t_{D})F^{\times}_{1}(\phi_{S0}-\frac{\pi}{4})],\end{align}
\end{subequations}
where $\mathcal{F}[\dots]$ denotes the Fourier transformation, $\zeta_{1}(f)=\phi_{S0}-\pi ft_{D}$, $\zeta_{2}(f)=\phi_{S0}+\pi ft_{D}$, $f_{0}\approx3.176\times10^{-6}{\rm Hz}$ is the orbital frequency of the TianQin satellites around the Earth, $\phi_{S0}$ is the initial location of sources, and $\widetilde{h}_{+,\times}(f)$ are Fourier transform of ${h}_{+,\times}(t)$.
For more details, please refer to Appendix \ref{sec:app2}. 
We emphasize that although we assume a circular orbit in Eq. (\ref{eq:h_t}), Eqs. (\ref{eq:h_t_response}) are applicable to general orbits, such as the eccentricity is not zero.

For the post-Newtonian waveform, it has been suggested that a waveform up to 2PN order is sufficiently accurate for the precise measurement of \ac{SBBH} with space-based \ac{GW} detectors \cite{Mangiagli:2018kpu}.
Therefore, we adopt the restricted 3PN waveform with and eccentricity for $\widetilde{h}_{+,\times}(f)$ throughout our calculation \cite{PhysRevD.52.2089,PhysRevD.80.084043,Feng:2019wgq}.
The choice of a higher order \ac{PN} waveform is to be conservative, especially considering the better sensitivity of TianQin in higher frequencies.
We ignore the spin of \acp{SBBH} as the effect is expected to be minor in low frequencies \cite{Nishizawa:2016jji}.

\subsection{Signal-to-noise ratio}\label{sec:SNR}

The recorded data $s(t)$ contains two parts: the noise $n(t)$ and the \ac{GW} signal $h(t)$,
\begin{equation}
  s(t)=h(t)+n(t).
\end{equation}
For the analysis of \ac{GW} signal, it is convenient to define the \emph {inner product} between two waveforms $h_{1}(t)$ and $h_{2}(t)$ \cite{Cutler:1994ys}
\begin{equation}\label{eq:in_prod}
  (h_{1}|h_{2})=4 \Re \int_{0}^{+\infty}{\rm d}f \,\frac{\widetilde{h}_{1}^{\ast}(f)\widetilde{h}_{2}(f)}{S_{n}(f)},
\end{equation}
where $\widetilde{h}_{1}(f)$ and $\widetilde{h}_{2}(f)$ are the Fourier transform of $h_{1}(t)$ and $h_{2}(t)$, respectively, and $^\ast$ represents complex conjugate, $S_{n}(f)$ is the power spectral density of detector noise $n(t)$.
The pure noise for TianQin is characterized with
\begin{eqnarray}\label{eq:PSD}
  S_N(f)          =\frac1{L^2}\left[\frac{4S_a}{(2\pi f)^4}\Big(1+\frac{10^{-4}{\rm Hz}}f\Big)+S_x\right],
\end{eqnarray}
with $S_a^{1/2}= 1\times 10^{-15}\Sa$, $S_x^{1/2} = 1\times 10^{-12}\Sx$ \cite{Luo:2015ght}.
However, in higher frequencies, when the low frequency approximation fails, the detector response to a given source falls rapidly.
Therefore, we instead use the effective sensitivity curve $S_n$ for the inner product in the actual calculation,
\begin{equation}
S_n(f)       =\frac{3}{20}\frac{S_N(f)}{\overline{R}(2\pi f)}
\end{equation}
where $\overline{R}(2\pi f) $ is the averaged response function, 
\begin{equation}
  \overline{R}(2\pi f) =\frac{3}{20}\times\frac{g(2\pi f\tau)}{1+0.6(2\pi f\tau)^2}
\end{equation}
where $\tau=L$ is the light travel time for a TianQin arm length, and the function $g(x)$ approaches unit for the lowest order approximation, with the higher order approximation listed in \cite{Wang:2019ryf}.

For a given detector, the optimal \ac{SNR} $\rho$ of a signal $h(t)$ is defined as the square root of inner product of $h$ with itself \cite{Cutler:1994ys}
\begin{equation}\label{eq:rho_opt}
  \rho = (h|h)^{1/2}=\sqrt{4 \Re \int_{0}^{+\infty}{\rm d}f \,\frac{\widetilde{h}^{\ast}(f)\widetilde{h}(f)}{S_{n}(f)}}.
\end{equation}
If multiple detectors observe the same event simultaneously, then the overall \ac{SNR} is defined as the root sum square of the individual \ac{SNR} for the $k$th detector $\rho_{k}$,
\begin{equation}
  \rho=\sqrt{\sum_{k}\rho_{k}^{2}}.
\end{equation}

The nominal configuration of the TianQin constellation as proposed in \cite{Luo:2015ght} follows a ``3 months on+3 months off'' observation pattern, causing gaps in the recorded data. 
As a result, the \ac{PN} waveform has to be set zero for certain range of frequencies in (\ref{eq:in_prod}).
The frequency boundaries can be found from the instantaneous frequency at the time $t$ before the merger time $t_c$,
\begin{equation}\label{eq:t2f}
  f=(5/256)^{3/8}\frac{1}{\pi} \mathcal{M}^{-5/8}(t_c-t)^{-3/8}\,,
\end{equation}
where the leading order of \ac{PN} expansion is used.
In practice, with a given merger time $t_c$, we perform cutoff on frequencies when the detector is not operating using Eq. (\ref{eq:t2f}) and then apply Eq. (\ref{eq:FT_response}) upon the truncated waveforms.

In this paper, we also consider the so-called twin constellation configuration of TianQin, which involves 2 three-satellite constellations perpendicular to each other while both being nearly perpendicular to the ecliptic plane.
The twin constellations could alleviate the data gap issue of the one constellation configuration through the relay of observation.

\subsection{Fisher information matrix}\label{sec:FIM}
For a \ac{GW} signal $h(t, \bm{\lambda})$, where the true physical parameter are $\bm{\lambda}$,
the contamination of noise in the data means that it is probable that the maximum likelihood parameter $\hat{\bm{\lambda}}$ would be shifted from the true parameter by $\Delta\bm{\lambda}$: $\hat{\bm{\lambda}}=\bm{\lambda}+\Delta\bm{\lambda}$.
The \ac{FIM} is a useful tool to assess the covariance matrix associated with the maximum likelihood estimate \cite{Cutler:1994ys},
\begin{equation}
  p(\Delta\bm{\lambda})\approx\sqrt{{\rm det}(\Gamma/2\pi)}\exp(-\frac{1}{2}\Gamma_{ij}\Delta\lambda^{i}\Delta\lambda^{j}),
\end{equation}
where $\Gamma_{ij}$ is the \ac{FIM},
\begin{equation}
  \Gamma_{ij}=\left(\frac{\partial h}{\partial\lambda^{i}}\middle|\frac{\partial h}{\partial\lambda^{j}}\right),
\end{equation}
and the Cramer-Rao bound of the covariance matrix $\Sigma$ is given by the inverse of $\Gamma$, $\Sigma=\Gamma^{-1}$.
For a network of detectors, the overall \ac{FIM} is the summation over component \acp{FIM},
\begin{equation}\label{eq:sumFIM}
  \Gamma_{ij}=\sum_{k}\Gamma_{ij}^{k}.
\end{equation}
Therefore, one can estimate the uncertainty as the square root of the corresponding diagonal component of $\Sigma$, 
\begin{equation}
  (\Delta\lambda^{i})_{{\rm rms}}=\sqrt{\Sigma_{ii}}\,.
\end{equation}
One exception is the precision on the sky localization $\Delta\bar{\Omega}_{S}$, which can be obtained by \cite{Berti:2004bd}
\begin{equation}
  \Delta\bar{\Omega}_{S}=2\pi|\sin\bar{\theta}_{S}|(\Sigma_{\bar{\theta}_{S}\bar{\theta}_{S}}\Sigma_{\bar{\phi}_{S}\bar{\phi}_{S}}-\Sigma_{\bar{\theta}_{S}\bar{\phi}_{S}}^{2})^{1/2}.
\end{equation}
Notice that \ac{FIM} is only an approximation on the statistical uncertainty; thus, it cannot give an assessment on systematic uncertainty.
Also, the approximation would generally fail in low \ac{SNR} scenarios \cite{Vallisneri_2008,Rodriguez_2013}.

\section{Results}\label{sec:result}

\subsection{Detection number}

We first study the expected detection number of \acp{SBBH}.
For each mass model, we generate 200 Monte Carlo simulations for the corresponding merger catalogs.
A preset detection threshold on \ac{SNR} would then be applied to identify the detectable sources.

For each catalog, we first determine the number of merger events by randomly drawing from the merger rate distribution.
Then, for each event, we randomize over all possible parameters, including the component masses, redshift, coalescence time, sky location and orbital angular momentum.
We choose $\bar{\phi}_{S}$ and $\bar{\phi}_{L}$ to be uniform in the range $[0, 2\pi]$, $\cos\bar{\theta}_{S}$ and $\cos\bar{\theta}_{L}$ to be uniform in the range $[-1, 1]$, and the spatial distribution is chosen to be uniform in the comoving volume.
We limit the distance of sources to $0<z<2$, and we use the standard $\Lambda{\rm CDM}$ cosmological model ($h=0.679, \Omega_{\Lambda}=0.694, \Omega_{M}=0.306$ \cite{Ade:2015xua}).
The coalescence time is evenly distributed in the comoving frame.
The choice of the redshift upper limit is to be large enough so that the most optimal configuration would not exceed a preset threshold.
As we will see in Fig. \ref{fig:zMq_distri}, binaries with larger $t_c$ have less possibility to be detected; therefore, we also set an upper limit on $t_c$ to a large enough value so that the possibility of detecting event with higher $t_c$ is negligible.
\footnote{Such an upper limit on $t_c$ is determined individually for different mass models, truncating on when no single event is detectable during $t_c-5$yr and $t_c$.}

The detection threshold for \ac{SNR} is chosen to be 5, 8, and 12.
The choice of 5/8 is consistent with ground-based \ac{GW} detection traditions and was widely used for a number of similar studies in the field \cite{Sesana:2016ljz, Sesana:2017vsj, Kyutoku:2016ppx,Nakamura:2016hna,Isoyama:2018rjb,Wong:2018uwb}.
By inhereting such threshold, we are allowed to make direct comparisons with relevant literatures. 
We stress, however, that the threshold 5 is quite optimistic and should only be meaningful when considering a network of \ac{GW} detectors.
It has been suggested in \cite{Moore:2019pke} that a search using template bank would require an \ac{SNR} threshold of $\sim15$ for LISA detection, and the threshold could be reduced to $\sim9$ through multidetector observation. 
We apply a similar procedure and calculate the expected \ac{SNR} threshold to be $\sim12$ for TianQin \footnote{Here we adopt the 3PN waveform which was later also used for event rate and parameter estimation calculation. This is different from \cite{Moore:2019pke}.}.
Therefore, for the following calculation, we stick with the choice of 5, 8, and 12 on the \ac{SNR} threshold.

In order to draw better informed conclusions on the capability of TianQin, we consider four different observation scenarios with the following combination or single detectors:
\begin{enumerate}
\item \emph {TQ} for the TianQin constellation.
\item \emph {TQ I+II} for the putative twin constellation configuration of TianQin.
\item \emph {TQ + LISA} for the joint observation of TianQin and a LISA type detector (hereafter shorten as LISA).
\item \emph {TQ I+II + LISA} for the joint observation of TianQin I+II and LISA.
\end{enumerate}
We assume 5 years of operation time for TianQin, and 4 years for LISA, and we assume the same starting time for all detectors.
Finally, we adopt \cite{Cornish:2018dyw, Berti:2004bd} for the LISA power spectral density and orbit.

\begin{figure}[h]
\centering
\includegraphics[width=1\textwidth]{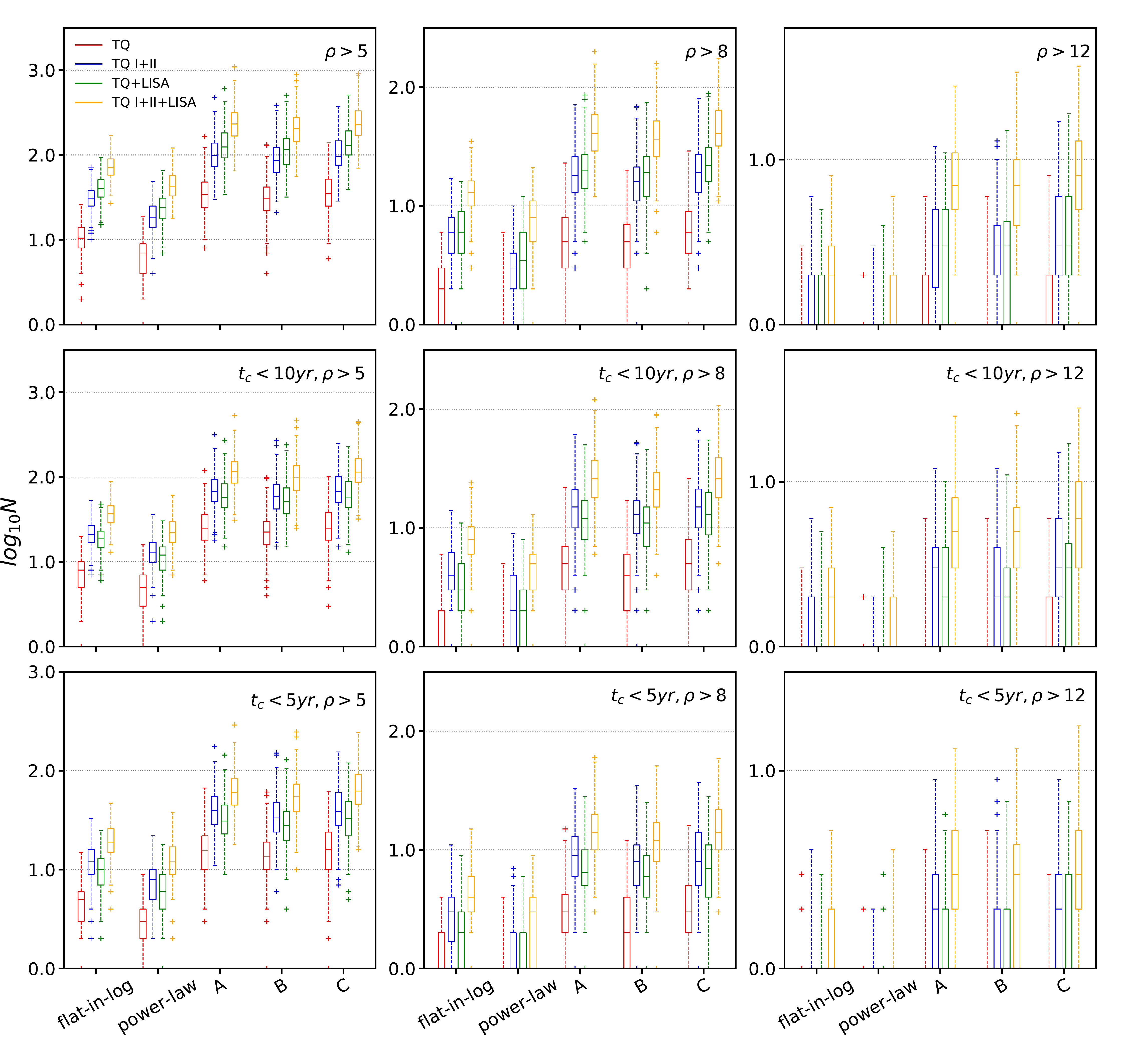}
  \caption{The box plots for detection number, under different mass distribution models (flat-in-log, power law, A, B, and C) and detector cases (TQ, TQ I+II, TQ+LISA, and TQ I+II + LISA).
The left column setting $\rho_{\rm thr}=5$, the middle column $\rho_{\rm thr}=8$ and the right column $\rho_{\rm thr}=12$.
  The top panel shows number for all events, while the middle panel for events that will merge in 10 years after the operation of TianQin, and the bottom panel for 5 years.}
\label{fig:det_num}
\end{figure}

The detection numbers over the whole mission lifetime for all detectors in all scenarios are shown as box plots in Fig. \ref{fig:det_num}.
For each case, a box plot illustrates the three quartiles with the middle line and the edges of the box; a whisker is used to indicate the extreme, or 1.5 times the box length when the furthest point is even further.
The top, middle and bottom panels correspond to the expected detection numbers for all events, for events merged within 10 years, and for events merged within 5 years, respectively.
The left, middle and right columns correspond to the detection threshold $\rho_{\rm thr}=5, 8, $ and 12, respectively.

For the pessimistic scenarios, i.e., adopting a threshold of $\rho_{\rm thr}=12$, TianQin is expected to detect at most order 1 \ac{SBBH}.
With a network of detectors, like TQ I+II, TQ+LISA and TQ I+II+LISA, the detection number is expected to increase, and order 1 of such binaries would merge within 5$-$10 years.
For the optimistic scenarios with $\rho_{\rm thr}=8$ for the detection threshold, adopting models A, B, and C for the mass models and considering all possible events, the expected detection number for TQ I+II, TQ+LISA, and TQ I+II+LISA could be a few dozens.
For each case, the 90\% credible interval spans 1 order of magnitude; for a given mass model, TQ I+II + LISA would have the most detections.

For a space-borne \ac{GW} detector like TianQin, the detection rate from \acp{SBBH} is mostly affected by two factors, the overall merger rate and the normalized mass distribution.
A more heavy-tailed \ac{SBBH} distribution (with larger portion of more massive \acp{SBH}) produces louder events for TianQin, while a higher merger rate leads to more events, and so both can lead to a larger detection rate.
We note that the mass distribution for the power-law model is significantly more heavy-headed (with larger portion of less massive \acp{SBH}) than all other models (Fig. \ref{fig:MassDistri}), while the merger rate of flat-in-log model is significantly lower than all other models (Fig. \ref{fig:MergerRate}).
As a result, the flat-in-log and power law models expect comparable detection numbers, which are consistently lower than those from models A, B, and C.

By adding more detectors, naturally more detections are expected.
This is reflected in Fig. \ref{fig:det_num} where green lines (TQ + LISA) and blue lines (TQ I+II) are always higher than red lines (TQ), while yellow lines (TQ I+II + LISA) are always the highest.
We note TQ I+II and TQ+LISA have comparable detection numbers. 
This is caused by the fact that TianQin has both sensitivity in the high frequency region but less observation time compared to LISA.

By comparing the left and right columns in Fig. \ref{fig:det_num}, one can see that a small difference in the \ac{SNR} threshold can make a big difference in terms of the detection numbers.
In the case $\rho_{\rm thr}=8$, almost all catalogs within all models predict a nonzero detection, with the most optimal cases predicting detection numbers to be reaching $\sim100$.
On the other hand, the case $\rho_{\rm thr}=12$ predicts much fewer detections.
The fact that the detection number is quite sensitive to the choice of detection threshold is consistent with the result of \cite{Moore:2019pke}, where a threshold of $\rho_{\rm thr}\sim15$ for LISA implies no detection at all.
We remark that efficient detection algorithms are needed for relatively weak signals.

There is a difference between the detection numbers in the top, middle, and bottom panels, but they are of comparable orders, with those in the middle panel being about 60\% of those in the top panel.
As is obvious from Eq. (\ref{eq:t2f}), the instantaneous frequency is sensitive to the amount of time left before the final merger, and the frequency evolution is slower for those far away from the merger than those close to the merger.
Limiting to sources that must merge within a certain amount of time will limit the frequency interval to be integrated in Eq. (\ref{eq:rho_opt}), hence leading to lower \ac{SNR} and smaller detection numbers.
Such fact indicates that for most mergers, a multiband \ac{GW} observation can be expected to be performed within a short time.

\subsection{Parameters estimation}

To study the precision of parameter estimation, we use the catalogs from the last subsection and focus on the {\it test events} that not only pass the detection threshold $\rho>8$ but also will merge within 5 years.
We use the same 3PN waveform, but allowing a nonzero value of eccentricity $e_0$, which is defined as the instantaneous eccentricity for when \ac{GW} frequency is 0.01Hz.
Since most sources have a minimum frequency of about $10^{-3}$Hz, we expected their eccentricities to be smaller than 0.1 \cite{Nishizawa:2016jji} due to \ac{GW} circularization.
We choose $e_{0}=0.01$ as a representative value  \cite{Nishizawa:2016jji}.

\begin{figure}[htbp]
\centering
\subfigure{
\includegraphics[width=0.3\textwidth, height=0.3\textwidth]{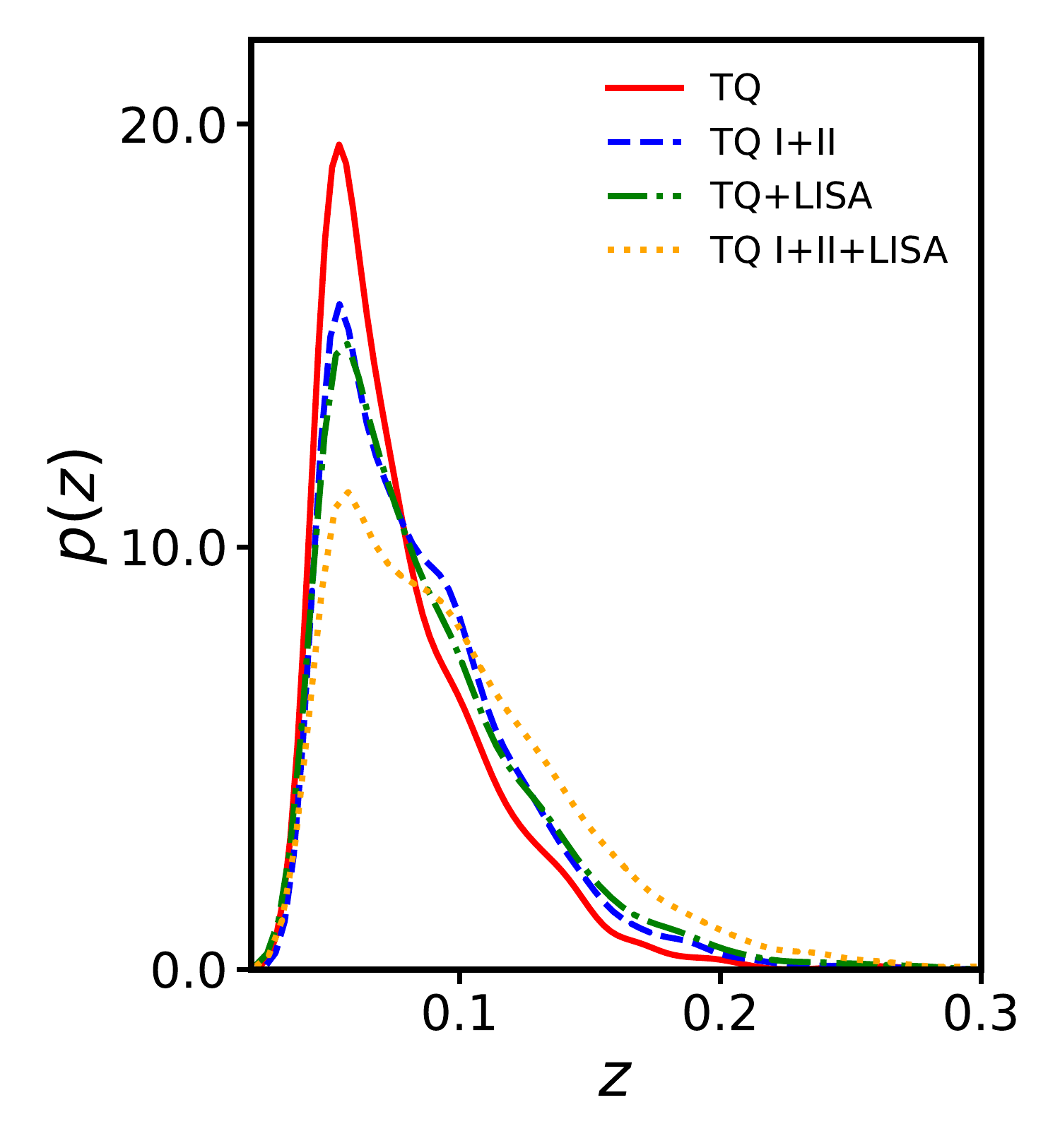}}
\subfigure{
\includegraphics[width=0.3\textwidth, height=0.3\textwidth]{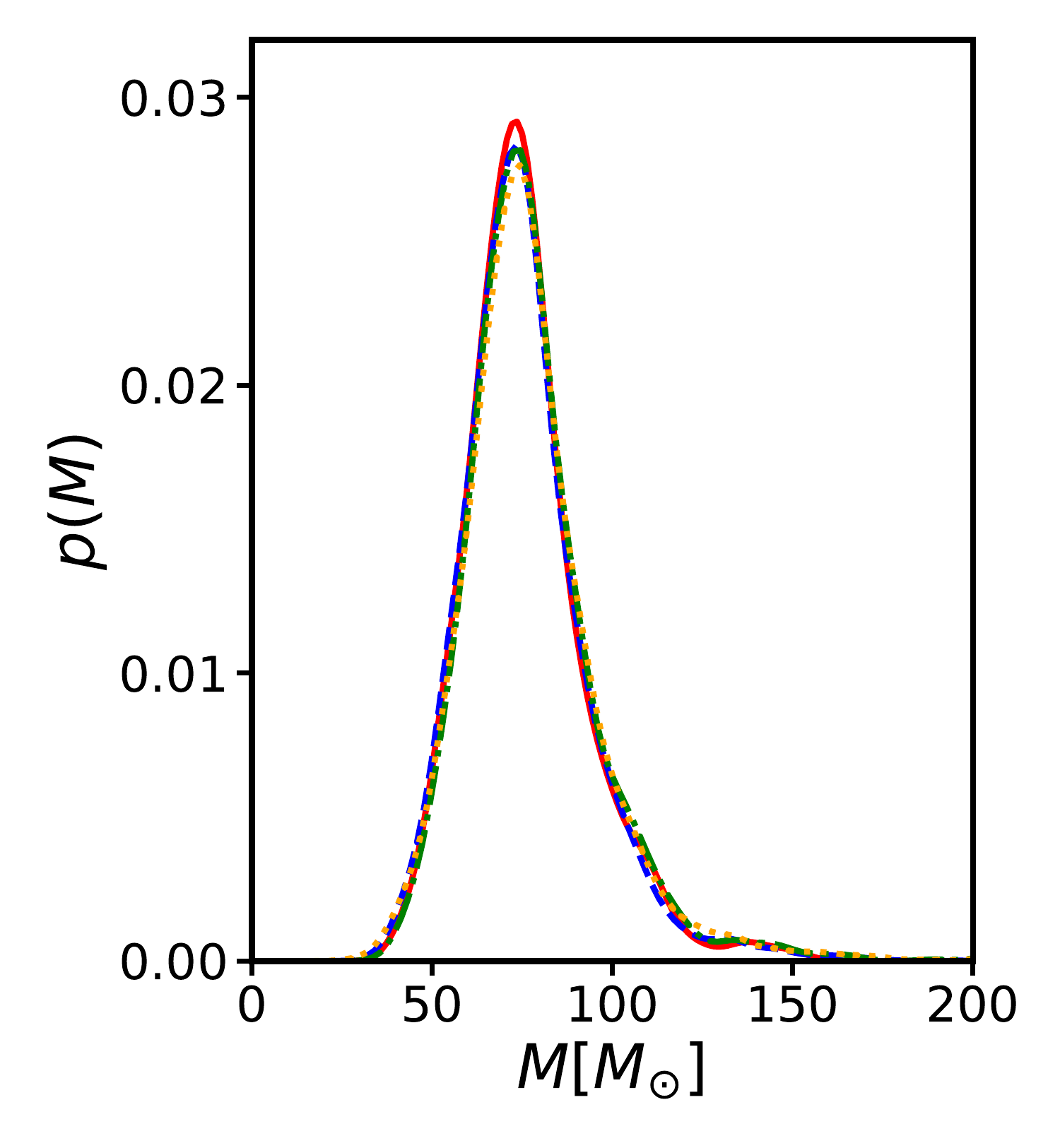}}
\subfigure{
\includegraphics[width=0.3\textwidth, height=0.3\textwidth]{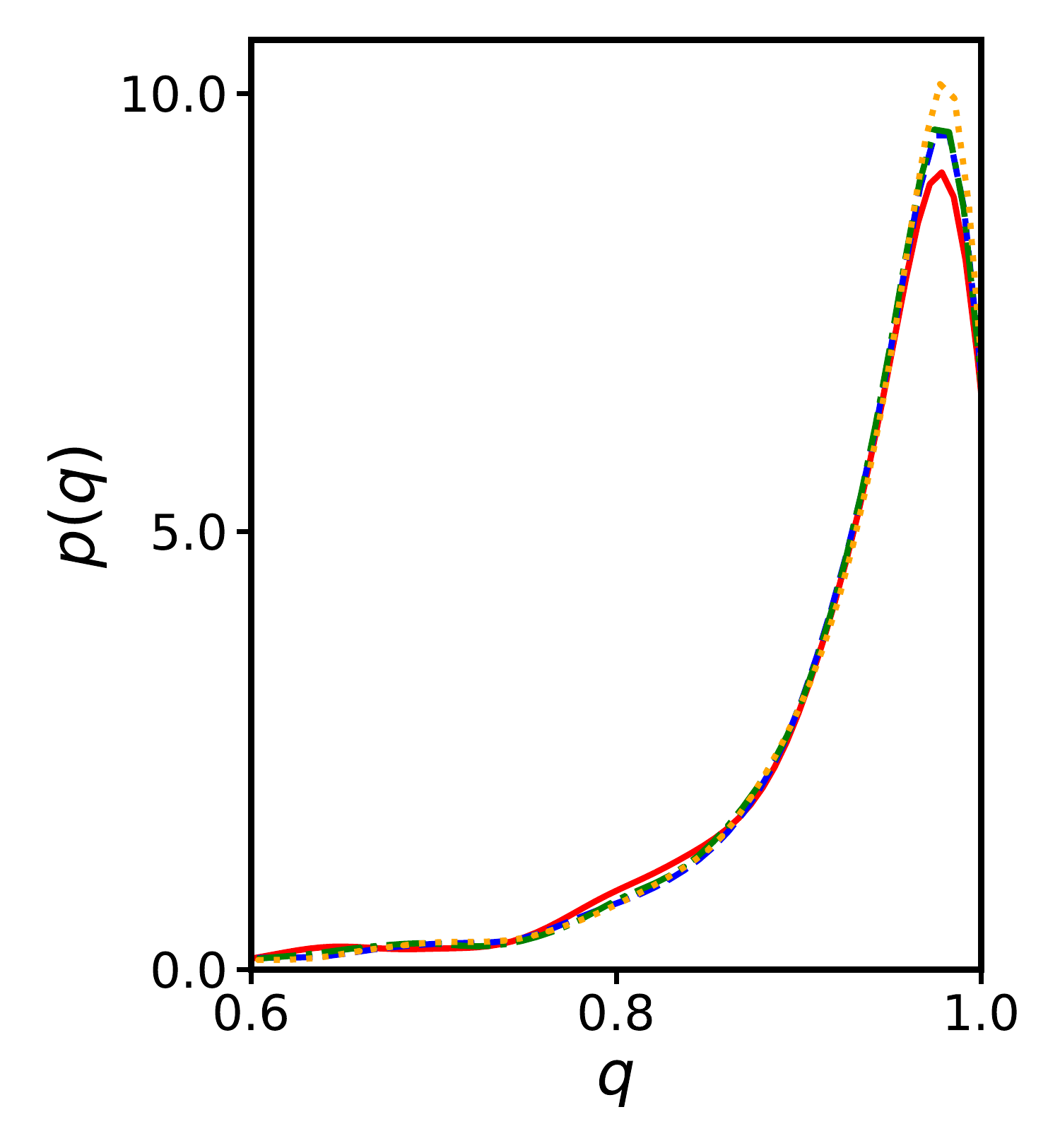}}
\caption{Parameters distributions for detectable events ($\rho >8$), assuming merge within 5 years (from left to right: redshift $z$, total mass $M$ and mass ratio $q$).}
\label{fig:zMq_distri}
\end{figure}

In Fig. \ref{fig:zMq_distri}, we present their distributions with respect to redshift $z$, total mass $M$ and symmetric mass ratio $\eta$.
We notice that the difference between different mass models is very minor,  and we show the result for model C as representatives.
Since we distribute the events uniformly in comoving volume, the events follow a dependence on luminosity distance $p(D_L)\propto D_L^2$, which is shown as rapidly rising below $z \sim 0.05$ in the redshift distribution.
However, events with too far a distance would hardly pass the detection threshold.
The two factors combined form a peak around $z \sim 0.05$ in the expected detected events.
For a space-borne \ac{GW} detector, all other factors being equal, a heavier SBBH always indicates a larger \ac{SNR}.
But the underlying distribution for \ac{SBH} mass falls for larger masses.
These two factors lead to a peak of $\sim 70 \msun$ for total mass $M$.
Finally, the mass ratio is heavily shifted toward unity; this is because the masses for \acp{SBH} have an upper limit, and an equally massive binary would form the heaviest binaries, meeting the preference of the detectors.
We also note that different detector combinations would only change the redshift distribution, as more detectors means higher \ac{SNR}, and the ability to detect more distant events.

\begin{figure}[htbp]
\centering
\subfigure[]{
\includegraphics[width=0.3\textwidth, height=0.3\textwidth]{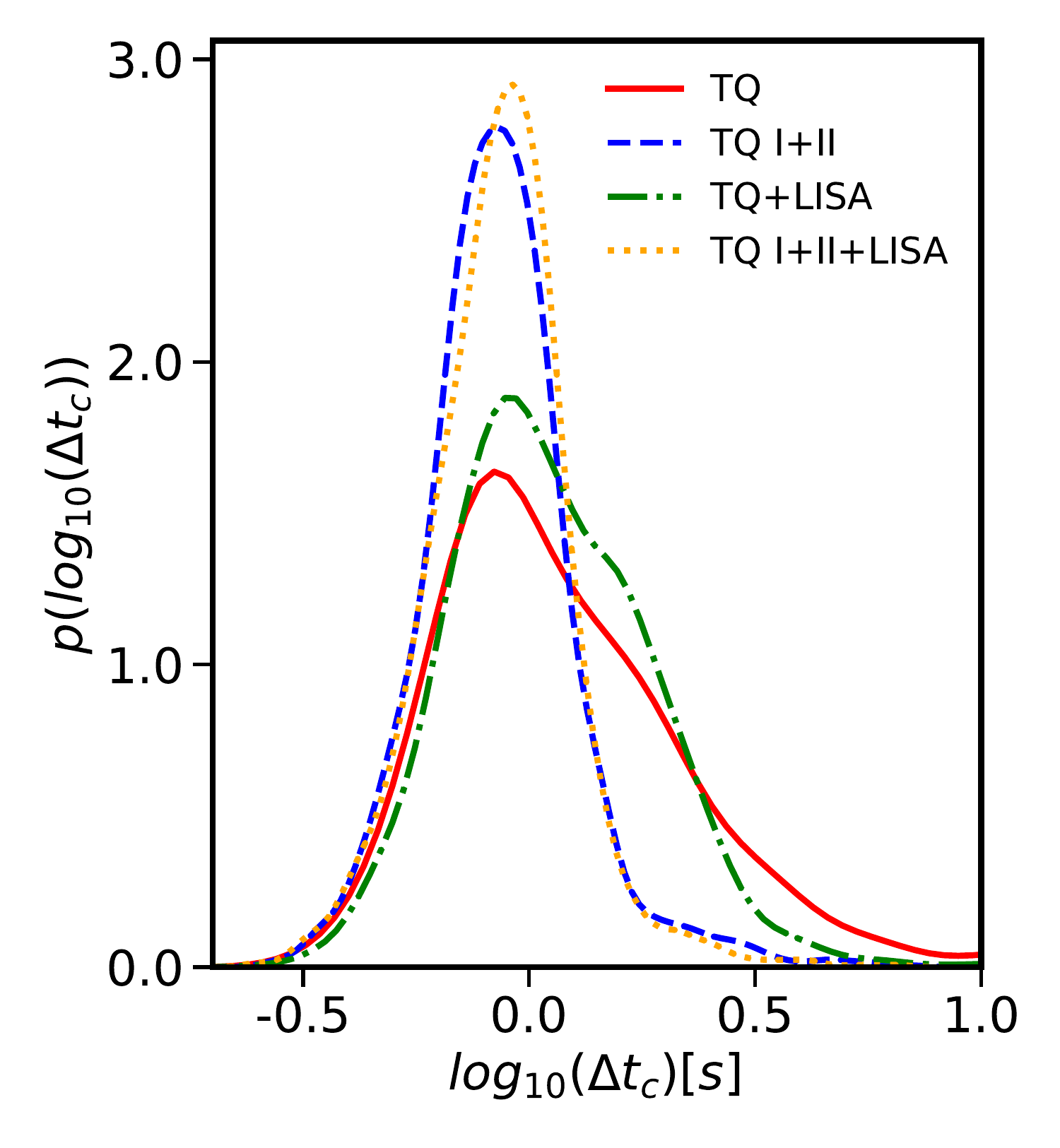}}
\subfigure[]{
\includegraphics[width=0.3\textwidth, height=0.3\textwidth]{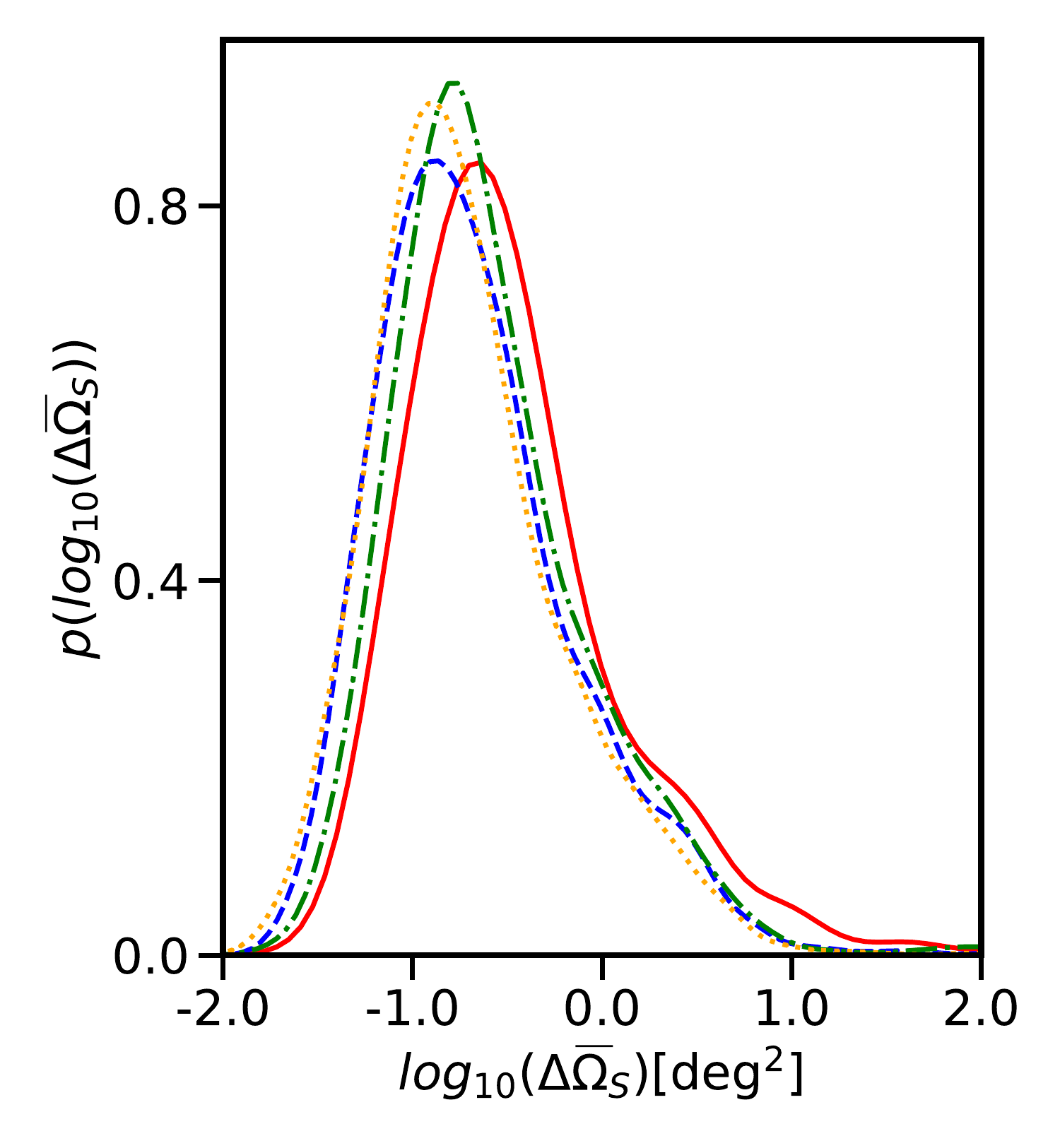}}
\subfigure[]{
\includegraphics[width=0.3\textwidth, height=0.3\textwidth]{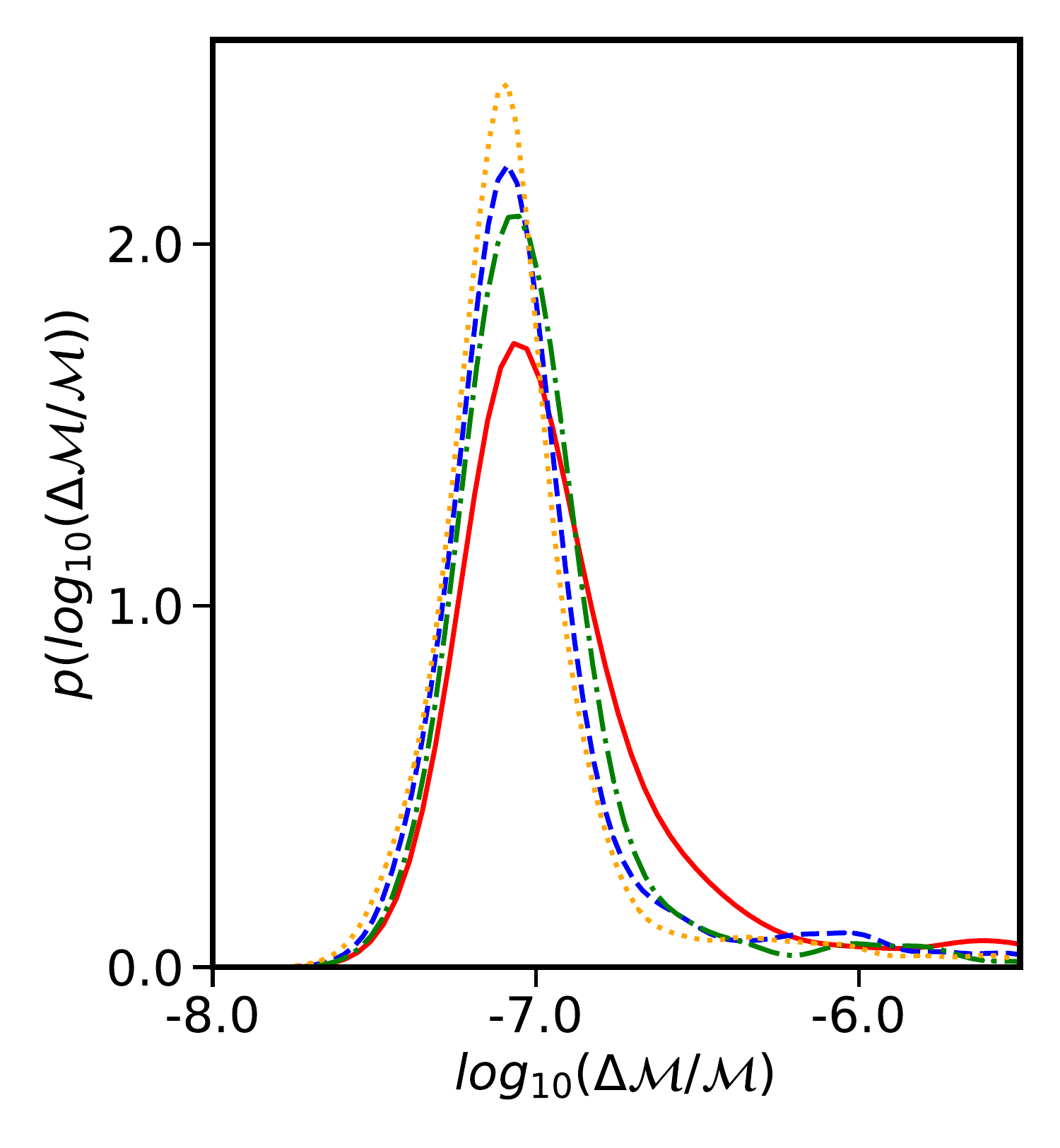}}
\subfigure[]{
\includegraphics[width=0.3\textwidth, height=0.3\textwidth]{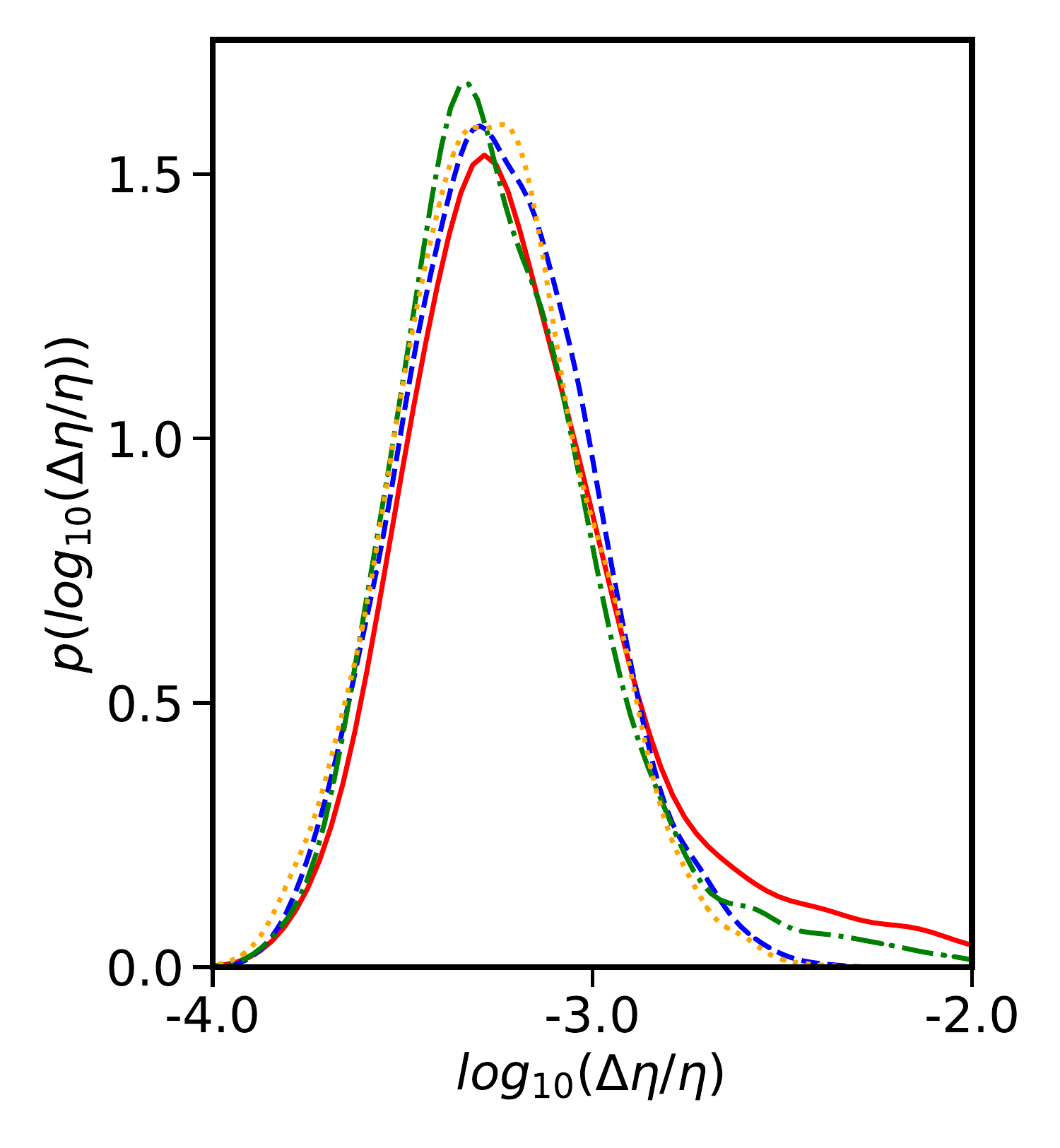}}
\subfigure[]{
\includegraphics[width=0.3\textwidth, height=0.3\textwidth]{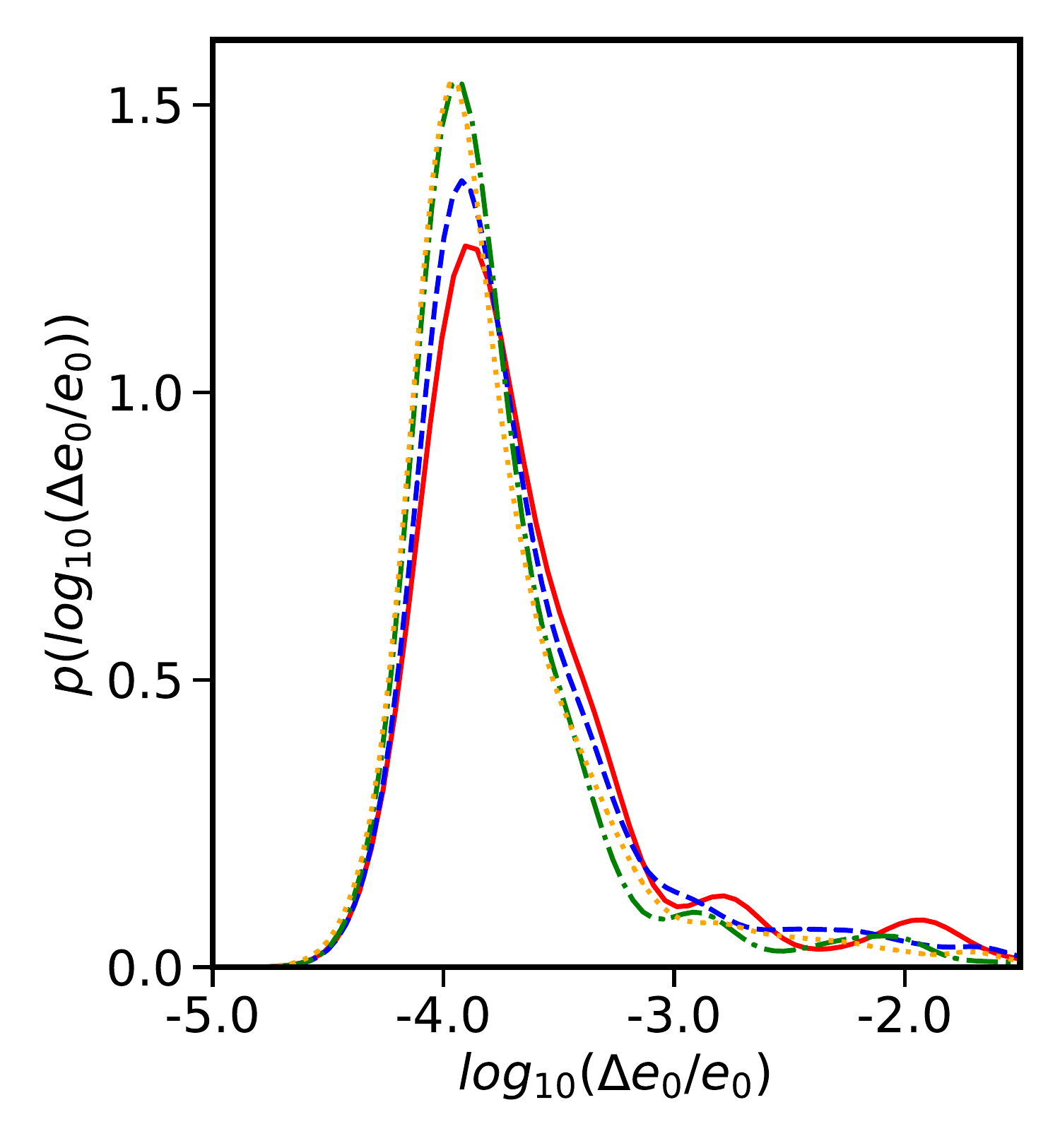}}
\subfigure[]{
\includegraphics[width=0.3\textwidth, height=0.3\textwidth]{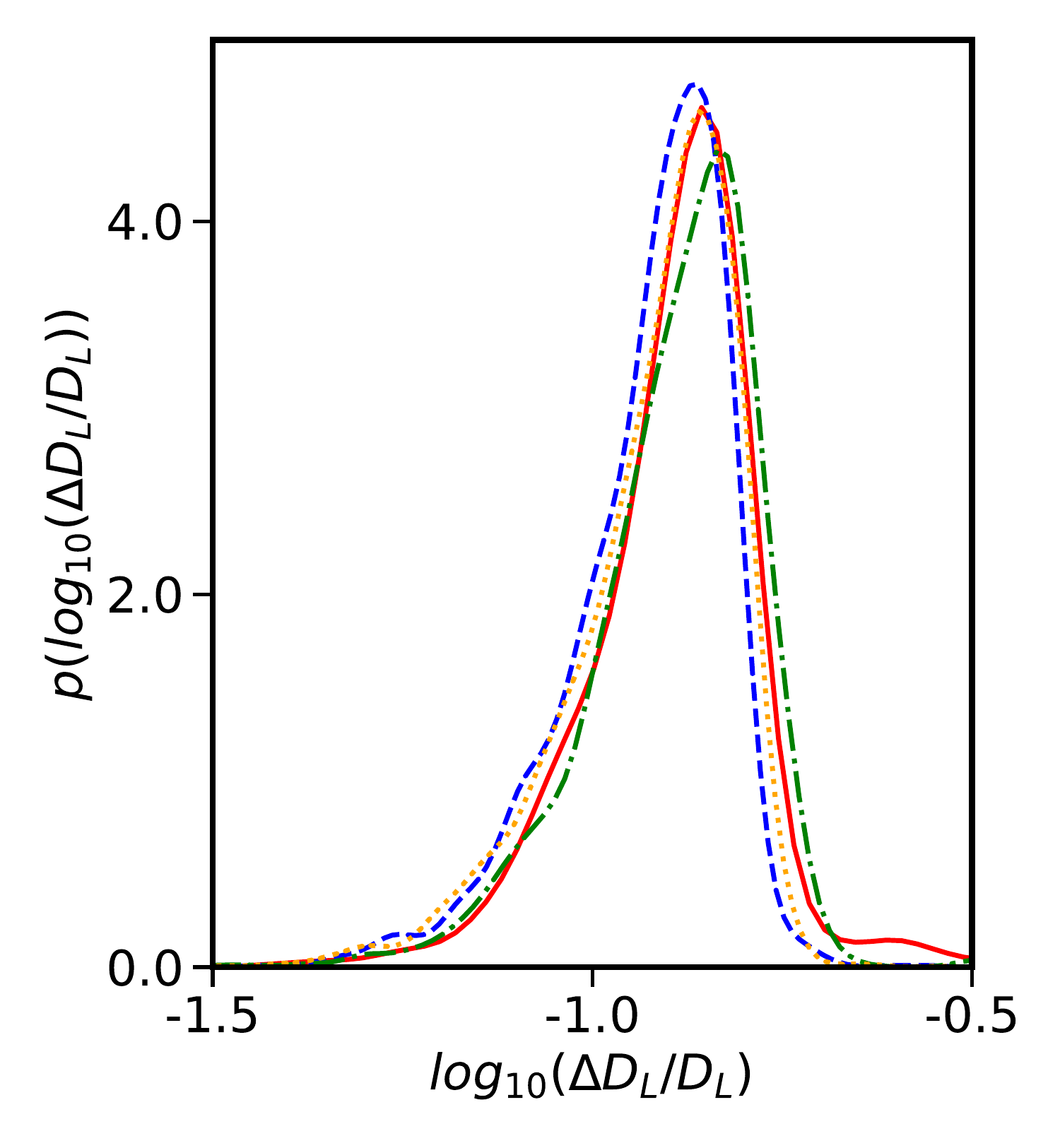}}
\caption{Estimation precision distributions of (a) coalescing time $t_{c}$, (b) sky localization $\bar{\Omega}_{S}$, (c) chirp mass $\mathcal{M}$, (d) symmetric mass ratio $\eta$ and (e) luminosity distance $D_{L}$.
Red solid line, blue dashed line, green dash-dotted line, and orange dotted line represent TianQin, TianQin twin constellations, TianQin+LISA, and TianQin twin constellations+LISA, respectively.}
\label{fig:PE_distri}
\end{figure}

We use the \ac{FIM} method to study the precision on the parameter set $\bm{\lambda}=\{t_{c}, \bar{\Omega}_{S}, \ln\mathcal{M}, \ln\eta,$ $ \ln e_{0}, \ln D_{L}\}$.
Since the difference caused by different mass models is small, we still use model C as an example.
The result is shown in Fig. \ref{fig:PE_distri}.
We notice that although the detection number differs quite a lot for different detector combinations, the normalized distribution for parameters is quite consistent for all parameters, and the spread of all parameter uncertainties are roughly about 1 order of magnitude, with those for the merger time and the luminosity distance being slightly narrower.
This is mainly due to the fact that the uncertainty in parameter estimation is roughly proportional to the inverse of \ac{SNR}, and applying a universal \ac{SNR} threshold leads to the very similar distribution.
More detectors mean larger numbers of high-\ac{SNR} detections, but it does not necessarily mean high-\ac{SNR} events have higher percentage.
That being said, including LISA in the detector network does seem to help reduce the tail with $\Delta t_{c}$, i.e. the uncertainty in the estimation of the merger time.

The expected uncertainties for localization are remarkable.
Using the most probable value from each plot as an indicator of TianQin's capability to measure the corresponding parameter, one can see that TianQin can predict the merger time with a precision of $\Delta t_{c}\sim 1$s, and report the sky location as precise as $\Delta \bar{\Omega}_{S} \sim 0.1$ deg$^2$.
This level of precision in space and time is good enough for \ac{EM} telescopes as well as for ground-based \ac{GW} observatories to prepare the examination toward the final merger moment.

Specifically, we remark the precise three-dimensional (3D) localization ability of TianQin, which is invaluable for multiband \ac{GW} observation as well as multimessenger observations, as it can greatly help in the identification of the host galaxy, which could open a bright possibility on \ac{GW} cosmology measurement.
Combined with a $\Delta D_{L}/D_{L}\sim $ 20\% relative error on luminosity distance, for a typical source located at redshift 0.05, the 3D error volume $\Delta V\sim D_{L}^{2}\Delta\bar{\Omega}_{s}\Delta D_{L}\sim$ 50Mpc$^{3}$. For the loudest events, $\Delta V$ could be as small as $\sim 2{\rm Mpc^{3}}$. Note that when the detection threshold increases, $\Delta V$ of the worst localized events would be improved \cite{DelPozzo:2018dpu}.
For an average number density of Milky-Way-like galaxy of 0.01 Mpc$^{-3}$, this means that one could pinpoint the host galaxy for the event \cite{Abadie:2010cf}.

The mass parameters are among the most precise parameters to be measured.
The chirp mass $\mathcal{M}$ has a huge effect on the phase of the \ac{PN} waveform; a slight change in $\mathcal{M}$ could lead to a huge dephase.
With a typical frequency of 0.01 Hz, and an observation duration of $\sim 10^8$s, an \ac{SBBH} is expected to rotate $\sim 10^6$ cycles during the observation.
Therefore, a relative error of $10^{-6}$ on frequency-related parameter can be expected, translating into a precise determination in the chirp mass relative error $\Delta \mathcal{M}/\mathcal{M} \sim 10^{-7}$.
The phase evolution depends also on the symmetric mass ratio $\eta$, but only on higher order terms, so the precision is much lower than chirp mass, but still can reach a remarkable 0.1\% relative error.

The eccentricity could also be very precisely determined, with the most probable relative uncertainty close to $\Delta e_0/e_0\sim 0.01\%$.
So even if the eccentricity $e_0$ is as small as 0.01 at 0.01 Hz, TianQin can still precisely measure it and use it as a promising tool to help unveil the formation mechanisms of \acp{SBBH}.

\begin{figure}[htbp]
\centering
\subfigure[]{
\includegraphics[width=0.3\textwidth, height=0.3\textwidth]{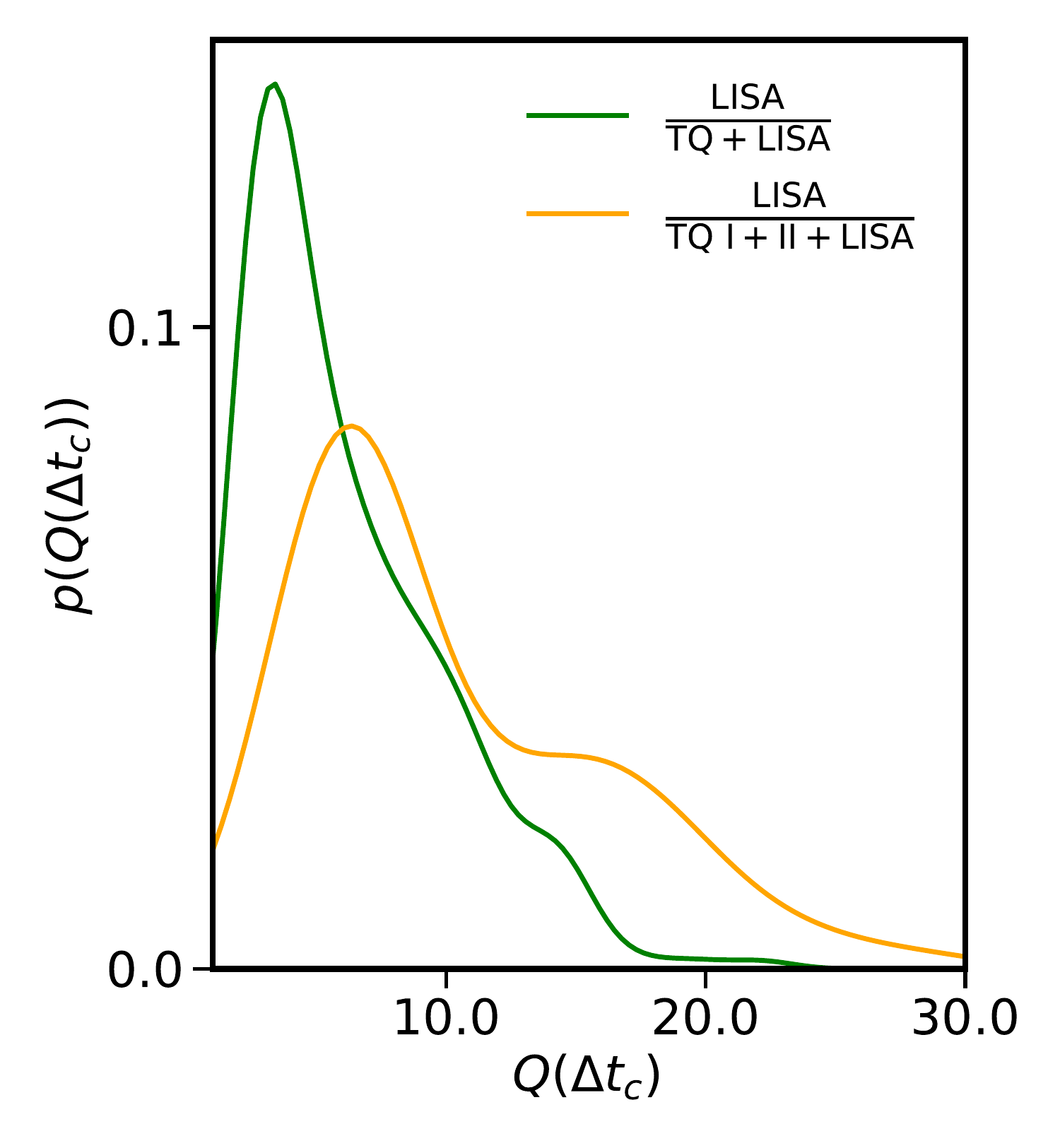}}
\subfigure[]{
\includegraphics[width=0.3\textwidth, height=0.3\textwidth]{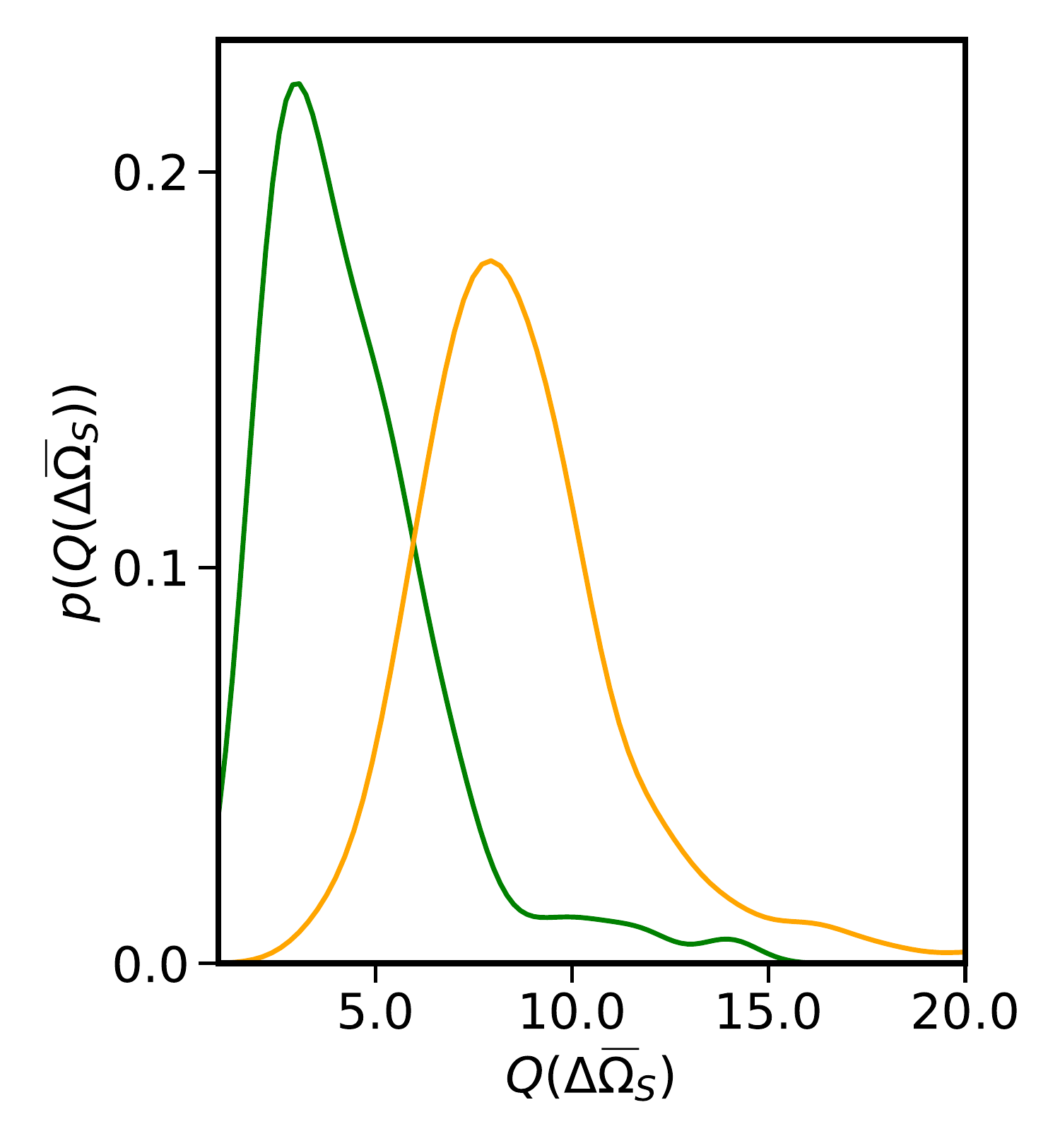}}
\subfigure[]{
\includegraphics[width=0.3\textwidth, height=0.3\textwidth]{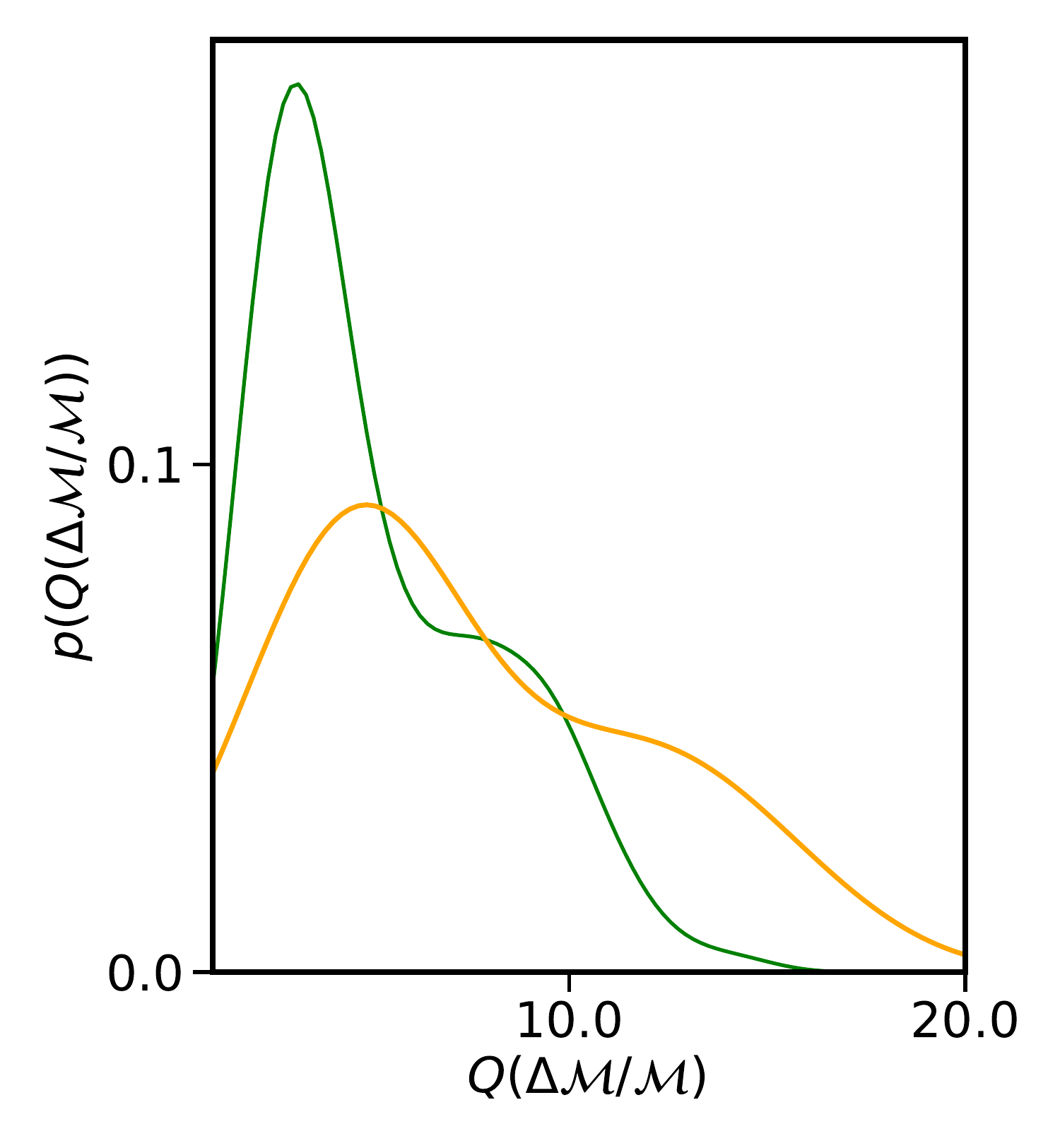}}
\subfigure[]{
\includegraphics[width=0.3\textwidth, height=0.3\textwidth]{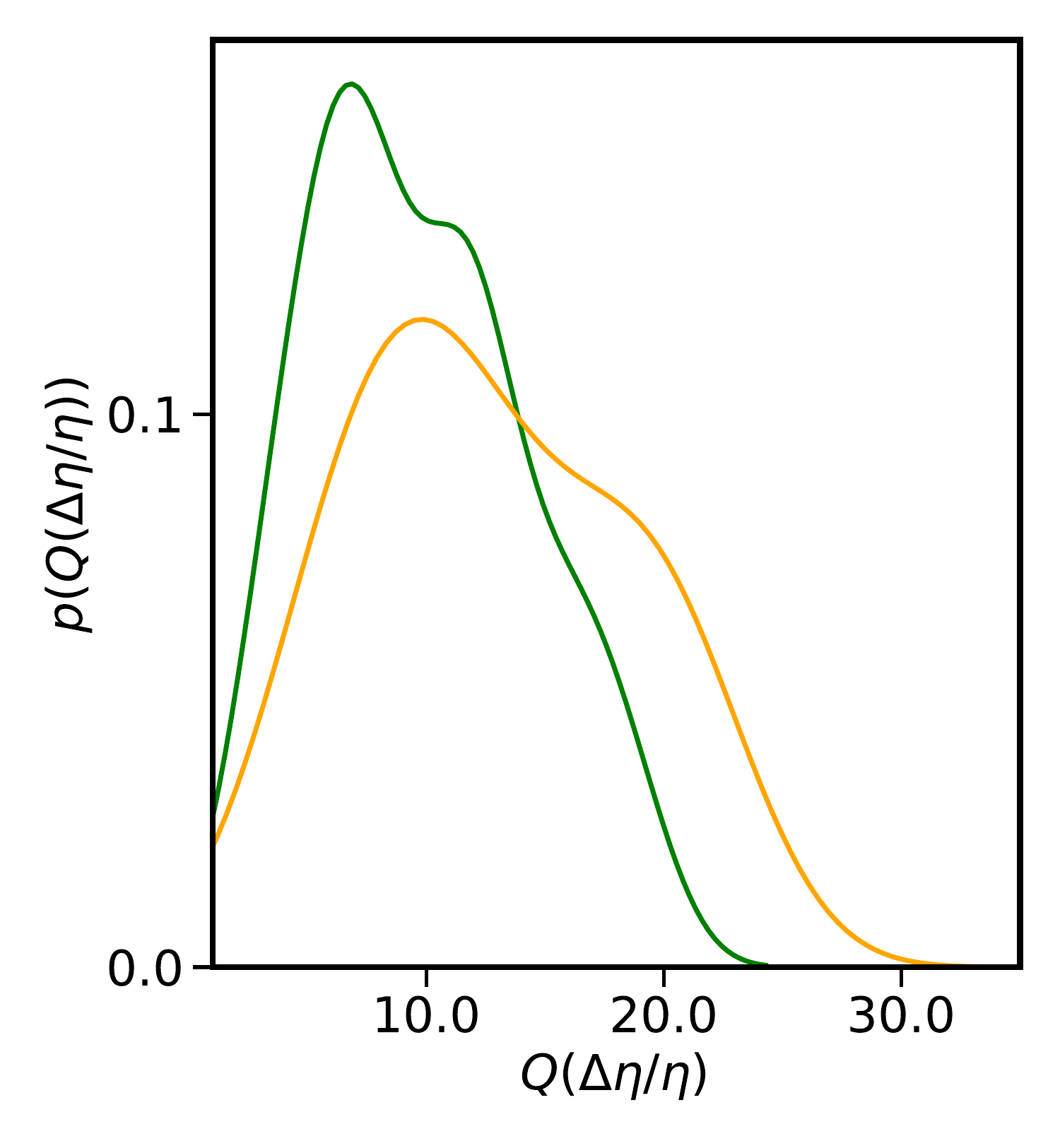}}
\subfigure[]{
\includegraphics[width=0.3\textwidth, height=0.3\textwidth]{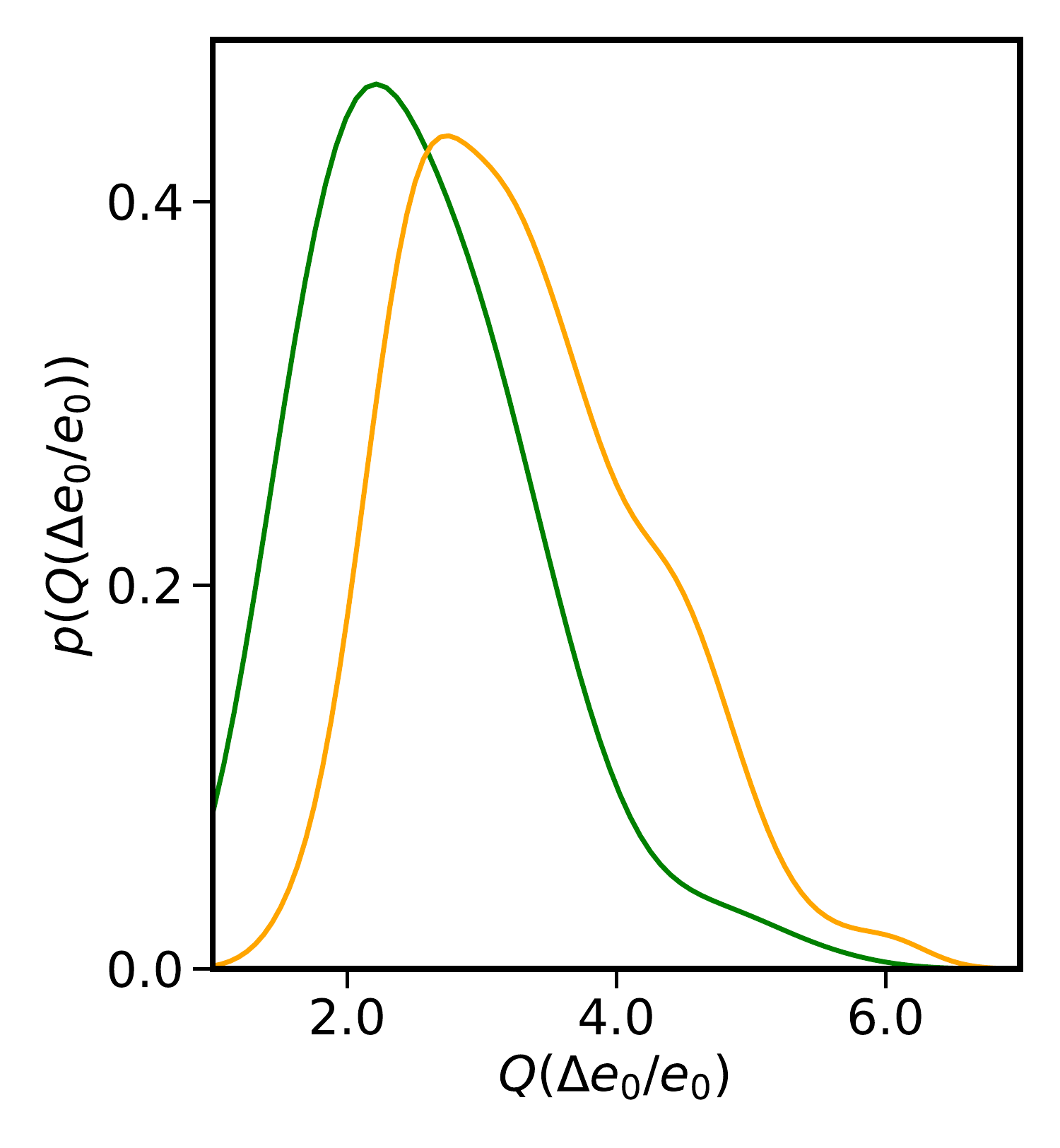}}
\subfigure[]{
\includegraphics[width=0.3\textwidth, height=0.3\textwidth]{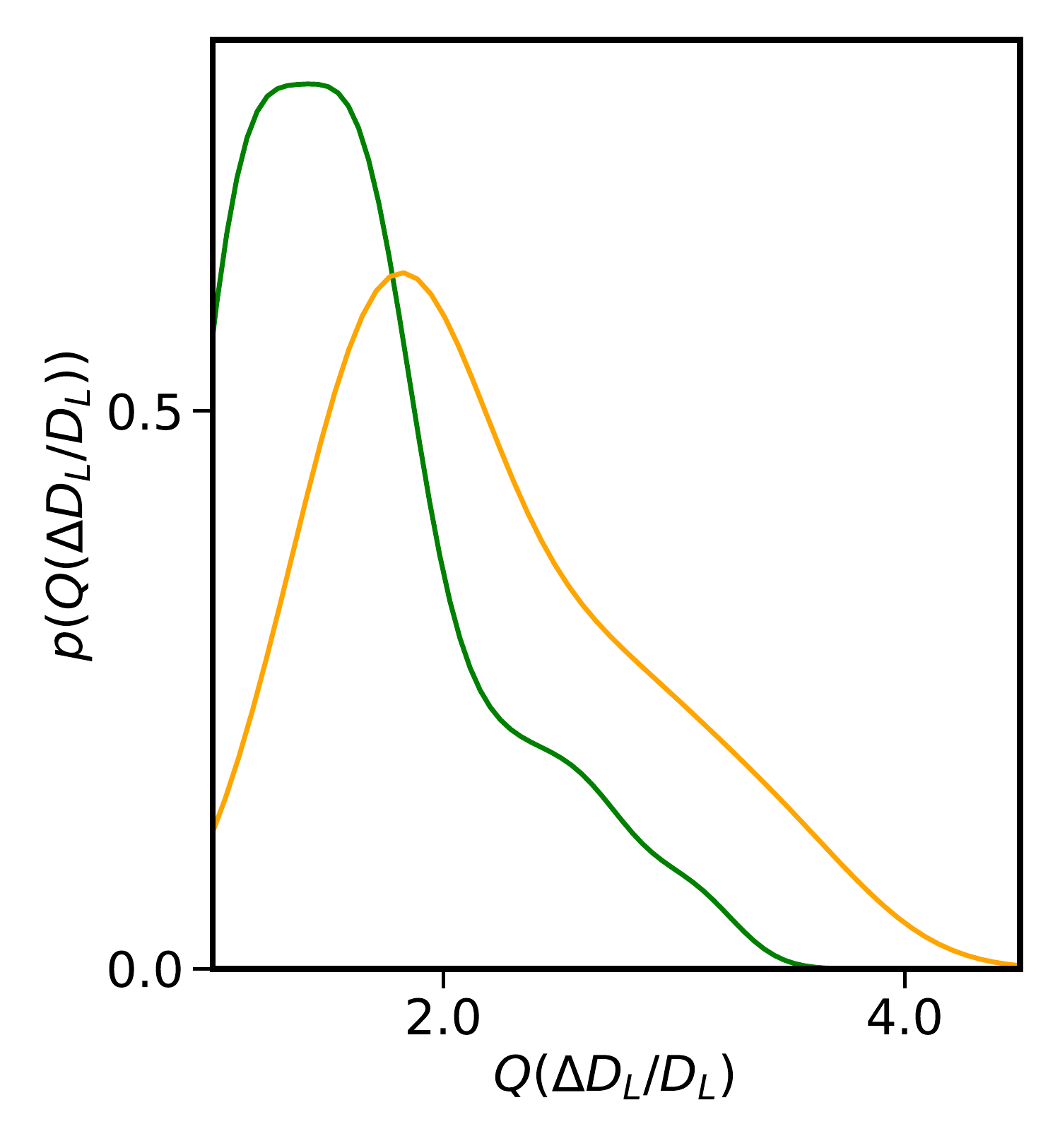}}
\caption{The improvement on the precision of parameter estimation of the (a) coalescing time $t_{c}$, (b) sky localization $\bar{\Omega}_{S}$, (c) chirp mass $\mathcal{M}$, (d) symmetric mass ratio $\eta$ and (e) eccentricity $e_0$, and (f) luminosity distance $D_{L}$,
for TQ+LISA (green) and TQ I+II + LISA (brown), with LISA being the reference.
}
\label{fig:PE_ratio_LISA}
\end{figure}

To see how TianQin can join a network of detectors to improve on the precision of parameter estimation for future space-based \ac{GW} missions, we use LISA as a reference mission and plot in Fig. \ref{fig:PE_ratio_LISA} the distribution of the ratio $Q$  between the precision of parameter estimation with LISA alone and the precision of parameter estimation by two detector networks involving TianQin: TQ+LISA (green line) and TQ I+II + LISA (orange line).
A larger value of $Q$ means a better improvement in precision.
One can see that the precision of the coalescing time $t_{c}$, the sky localization $\bar{\Omega}_{S}$, the chirp mass $\mathcal{M}$, and symmetric mass ratio $\eta$ can all be significantly improved, and for some of them the improvement can be close to 1 order of magnitude.
For parameters like the luminosity distance $D_L$, however, the improvement is less than 2, comparable to the improvement on the \ac{SNR}.
The reason is that the luminosity distance only affects the magnitude of \ac{GW}, which is often measured with less accuracy than the \ac{GW} phase.

\begin{figure}[htbp]
\centering
\subfigure[]{
\includegraphics[width=0.3\textwidth, height=0.25\textwidth]{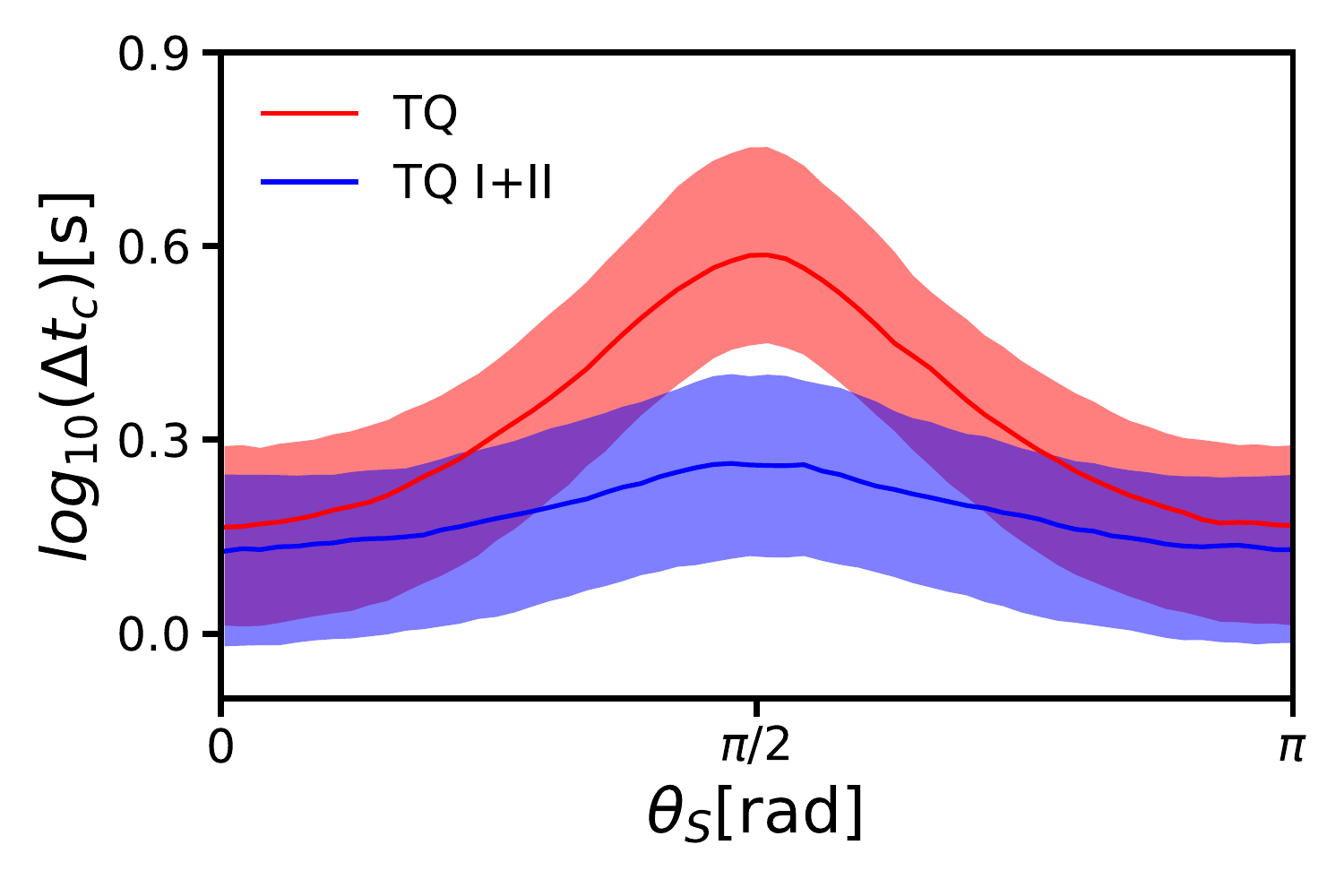}}
\subfigure[]{
\includegraphics[width=0.3\textwidth, height=0.25\textwidth]{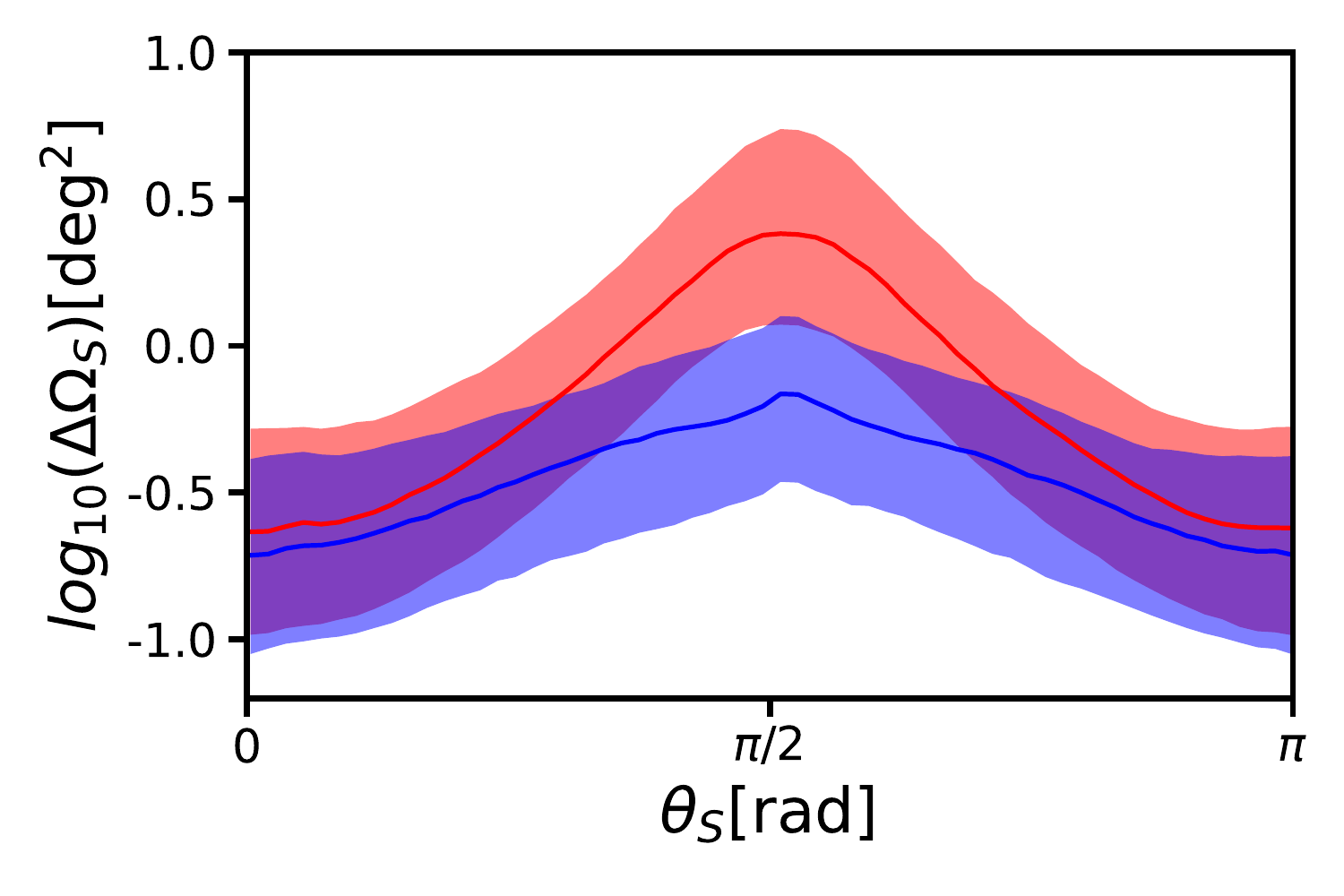}}
\subfigure[]{
\includegraphics[width=0.3\textwidth, height=0.25\textwidth]{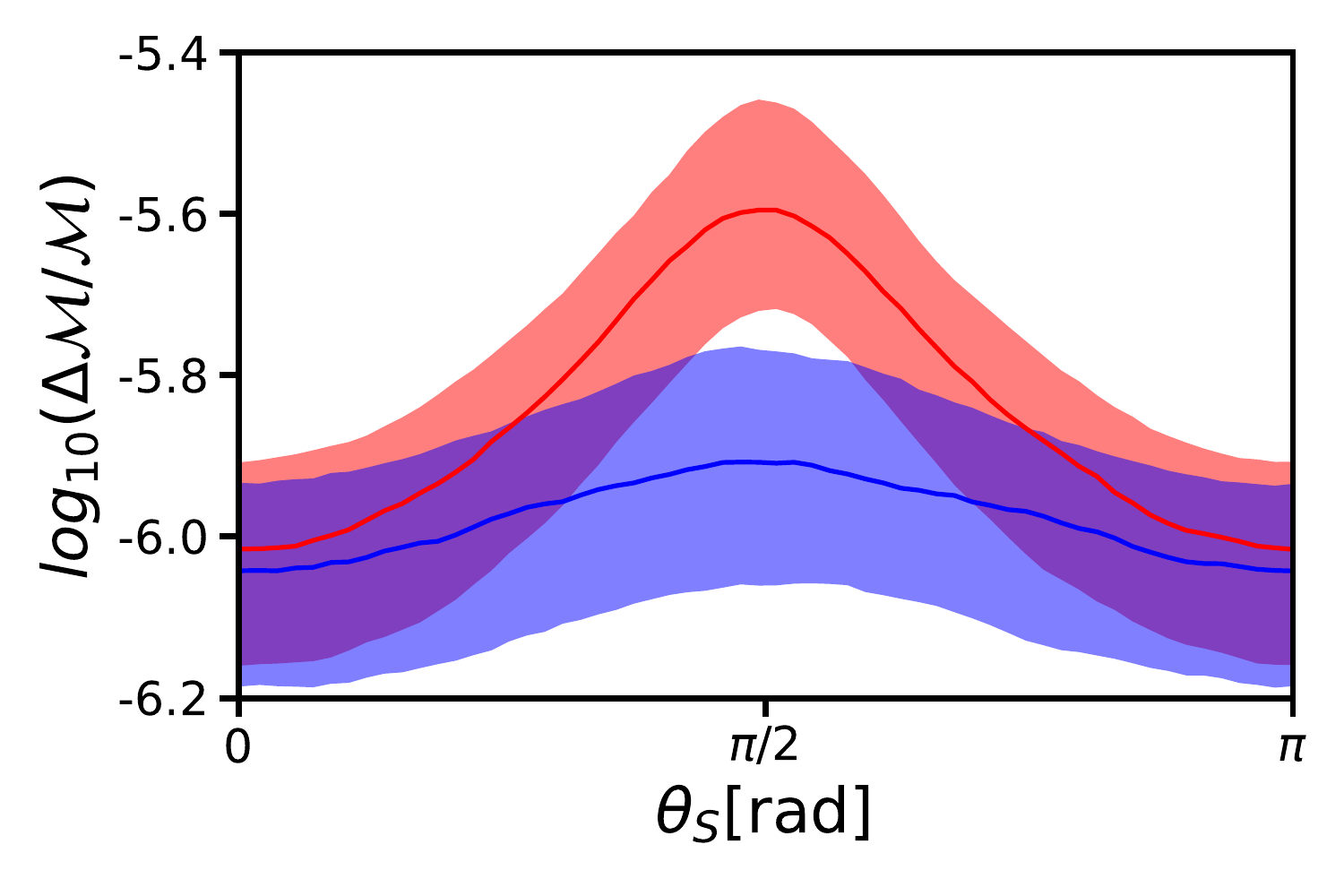}}
\subfigure[]{
\includegraphics[width=0.3\textwidth, height=0.25\textwidth]{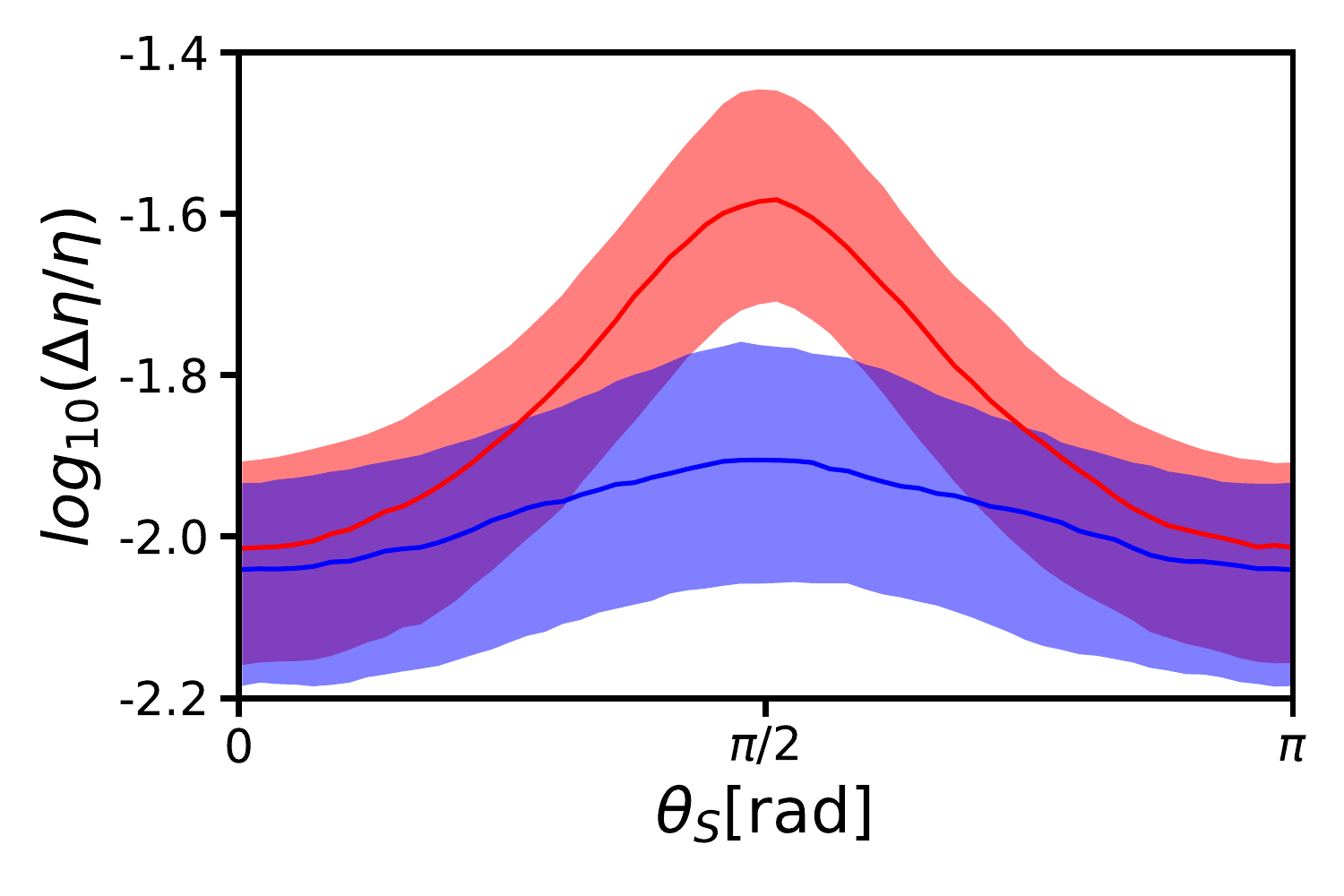}}
\subfigure[]{
\includegraphics[width=0.3\textwidth, height=0.25\textwidth]{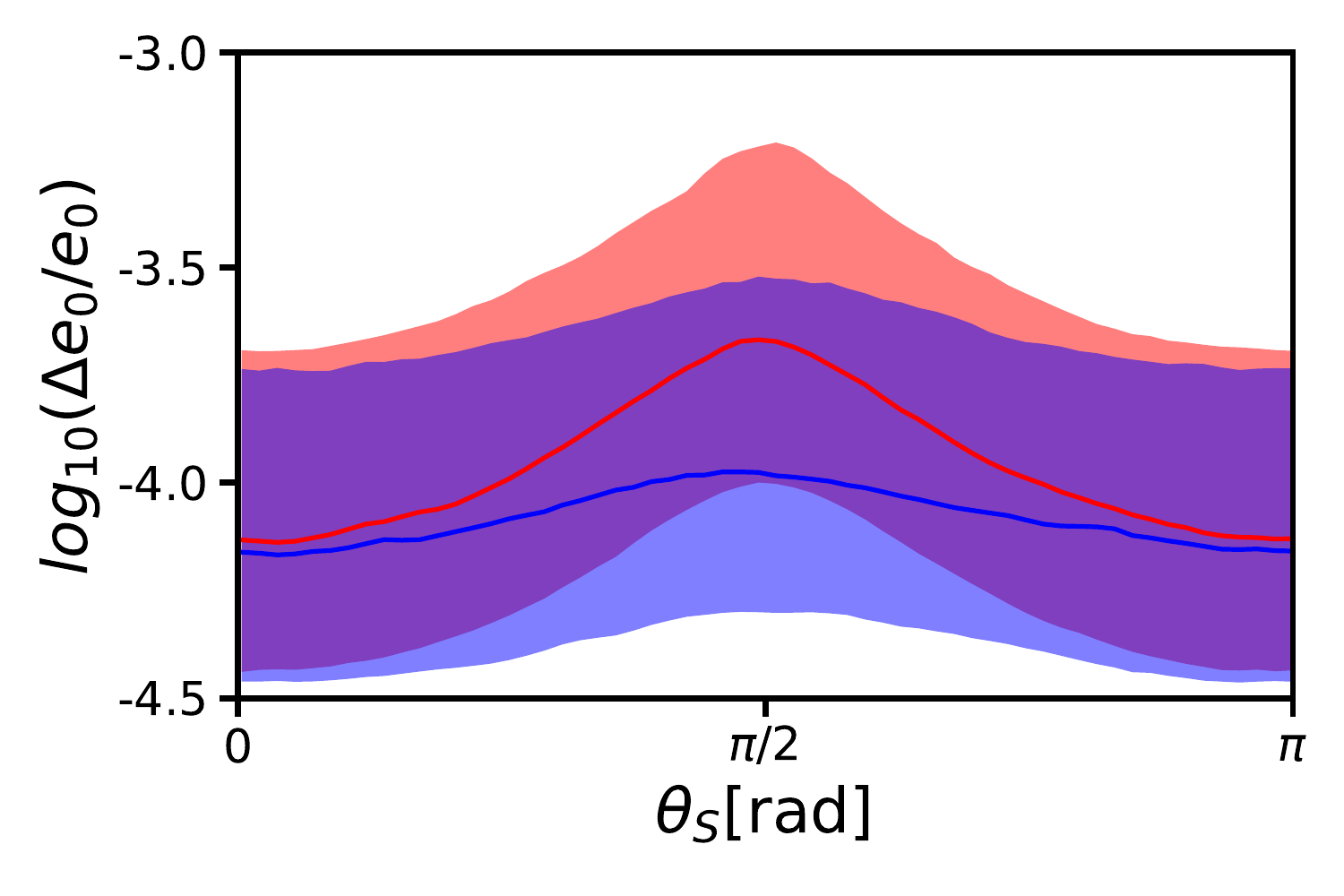}}
\subfigure[]{
\includegraphics[width=0.3\textwidth, height=0.244\textwidth]{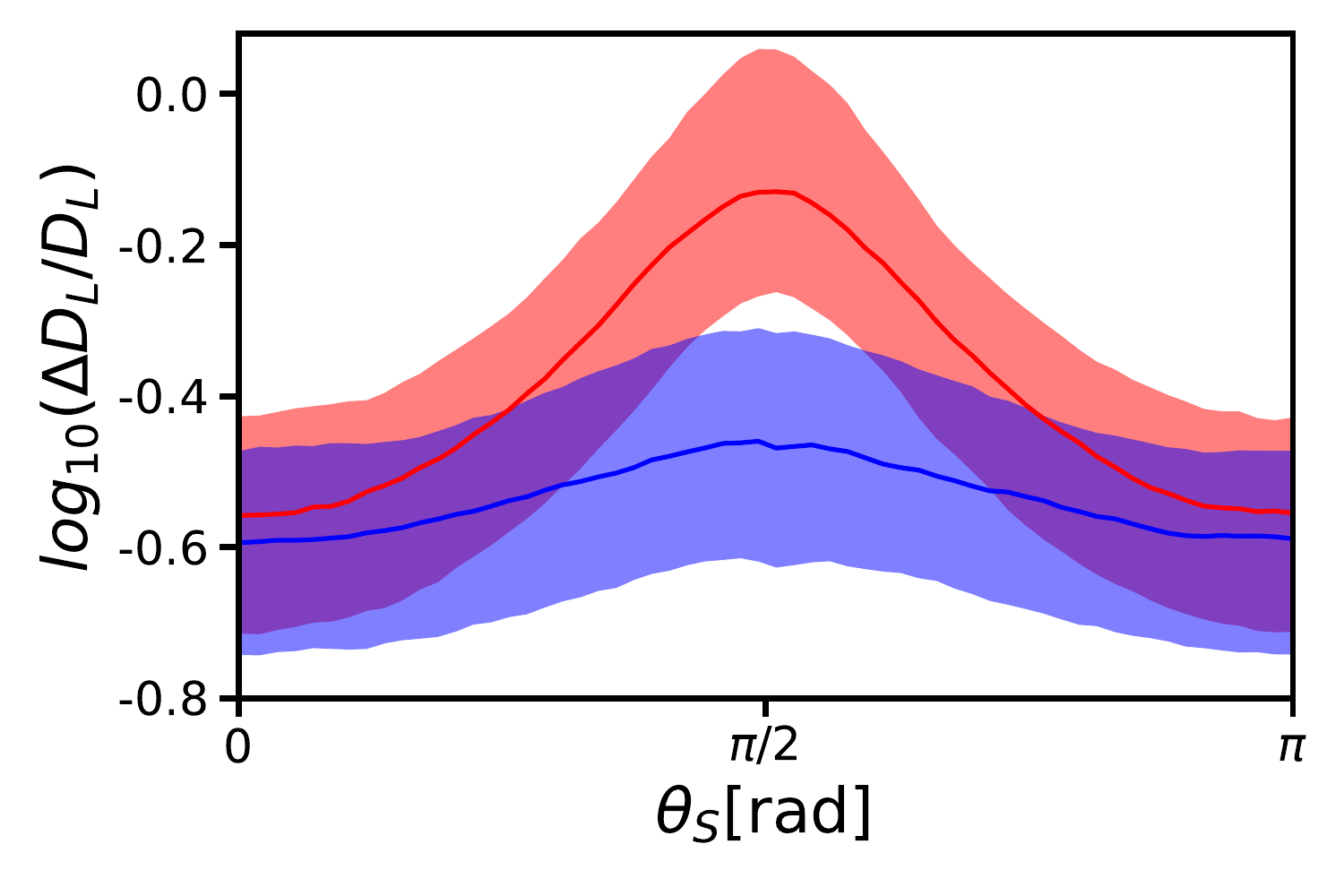}}
\caption{Parameter estimation precision dependence on altitude $\theta$ in the detector coordinate system, showing in (a) coalescing time $t_{c}$, (b) sky localization $\Omega_{S}$, (c) chirp mass $\mathcal{M}$, (d) symmetric mass ratio $\eta$, (e) eccentricity $e_{0}$, and (f) luminosity distance $D_{L}$.
The line being the median while the shaded region reflects the central 50\% credible interval.
}
\label{fig:PE_altitude}
\end{figure}

The orientation of the TianQin orbital plane is nearly fixed in space.
We want to know how this feature will affect the precision of parameter estimation for sources located at different directions in space.
For this purpose, we adopt a detector-based spherical coordinate system that uses the TianQin orbital plane as its equator.
In this coordinate system, a celestial object would have a constantly changing azimuth, but a fixed altitude.
We randomly choose $\phi_{S}$ and $\phi_{L}$ from the distribution $U(0, 2\pi)$, $\cos\theta_{L}$ from the distribution $U(-1, 1)$ and $t_{c}$ from  $U(0, 5)$ years for a group of \acp{SBBH} with $m_{1}=m_{2}=30\msun$ at $D_{L}$=200Mpc, and look at how the precision of parameter estimation varies with the altitude $\theta_S$.
The result is shown in Fig. \ref{fig:PE_altitude}.
One can see that precision for sources near the zenith and the nadir, corresponding to $\theta_S=0$ and $\theta_S=\pi$, is always better than that for sources near the equator, in the amount of about half to one decade.
This is consistent with the general expectation that a \ac{GW} detector has better sensitivity for sources near the zenith and nadir than for those near the equator.
In the putative TianQin I+II network, a new constellation orthogonal to the initial TianQin constellation is introduced and the two constellations operate consecutively and repeatedly.
We see in Fig. \ref{fig:PE_altitude} that TianQin I+II has improved all sky response as expected.

\begin{figure}[htbp]
\centering
\includegraphics[width=0.5\textwidth, height=0.5\textwidth]{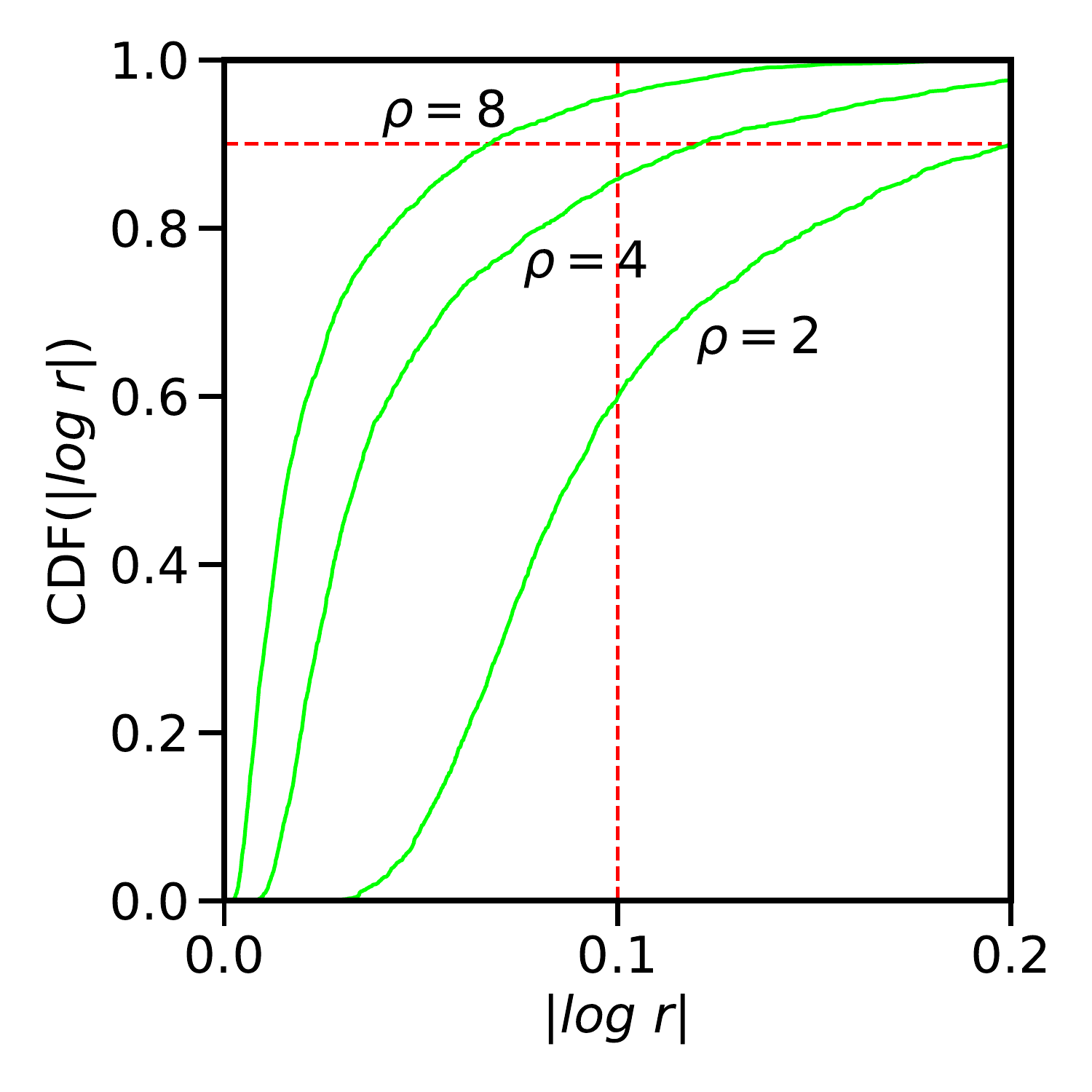}
  \caption{Cumulative distribution of logarithm of mismatch ratio $\log r$ for different \acp{SNR}, assuming the mass parameters of GW150914, to assess the validity of \ac{FIM}.  The curve of \ac{SNR} 8 is higher than the 90\% and $\log r=0.1$ point, meaning that the derived likelihood from \ac{FIM} and the exact value are close enough and the results from \ac{FIM} are trustworthy for \ac{SNR} as low as 8.}
\label{fig:log_r_abs}
\end{figure}

The aforementioned calculations are all based on the method of \ac{FIM}.
As mentioned in Sec. \ref{sec:FIM}, it is expected that the validity of \ac{FIM} will fade in low \ac{SNR} cases.
Therefore, we aim to investigate that to what extent can we trust the parameter estimation results from \ac{FIM}.
We follow Vallisneri to present the cumulative distribution for mismatch ratio $r$ over the isoprobability surface deduced from \ac{FIM} method \cite{Vallisneri_2008}.
The mismatch ratio $r$ quantifies the difference from the exact value of likelihood and the derived value approximated by the \ac{FIM} method.
By adopting the mass parameter of GW150914, we present the cumulative distribution of logarithm of $r$ for different \acp{SNR} in Fig. \ref{fig:log_r_abs}. 
We notice that for events with an \ac{SNR} of 8, for more than 90\% of the randomly drawn points from the isoprobability surface, their actual likelihood deviates only slightly from the derived value, marking the validity of \ac{FIM} in such \ac{SNR} level.
Furthermore, \ac{FIM} conclusions can be largely trusted for events with \acp{SNR} as low as 4.  

\section{Summary}\label{sec:summary}

In this work, we carry out a systematic study on the capability of TianQin in terms of observing \ac{SBBH}. 
We estimate the detection number of SBBHs as well as the precision of source parameter estimation with TianQin.
In order to make the result as robust as possible, we use five models for the mass distribution and the corresponding merger rate of \ac{SBBH}, i.e. the models flat-in-log, power law, A, B and C. 
In order to draw better informed conclusions on the capability of TianQin, we not only consider the detection capability of TianQin alone but also explore the detection capability of three detector networks containing TianQin: TQ I+II, TQ + LISA and TQ I+II + LISA.

We find that a network of multiple detectors is needed for the detection of \acp{SBBH}, for the pessimistic mass models (flat-in-log and power law) and the more strict \ac{SNR} threshold, $\rho_{\rm thr}=12$. 
With the more optimistic mass models A, B and C, TianQin is expected to detect a few \acp{SBBH}. 
What's more, if TianQin forms a detector network, such as TQ I+II, TQ + LISA and TQ I+II + LISA, the upper end of total expected detection number can reach over 10. 
When the SNR threshold is 8, the network of detectors is expected to detect $\sim100$ at most.

Using the \ac{FIM} method, we find that source parameters for the detected \acp{SBBH} events  can be precisely determined.
Using the most probable value from each plot in Fig. \ref{fig:PE_distri} as an indicator of TianQin's capability to measure the corresponding parameter, we find that TianQin can measure the chirp mass to the order $10^{-7}$, measure the symmetric mass ratio $\eta$ better than the order $10^{-3}$, forecast the merger time $t_c$ with a precision of the order $\sim $1s, determine the sky location of the source with a precision of the order $\sim 0.1$ deg$^2$, determine the luminosity distance to $20\%$ level, and measure the eccentricity $e_0$ to the order $10^{-4}$. 
The high precision in the determination of the source parameters is of great importance for many scientific purposes. 
For example, a precise prediction for the final merger moment is important for the follow-up multimessenger observation using \ac{EM} facilities and multiband \ac{GW} observation with ground-based \ac{GW} observatories; a high precision in the measurement of the eccentricity $e_0$ could help distinguish the formation channels of \acp{SBBH} and so on. 
A validity check for the \ac{FIM} method is performed, and the corresponding conclusion is trustworthy for signals with an \ac{SNR} as low as 8.

We highlight that the typical source localization error box has volume of the order $\Delta V\sim$ 50Mpc$^{3}$, which is so small that it contains only one Milky-Way-like galaxy in average, and this could greatly help in the identification of the host galaxy and make possible a great deal of science \cite{Fan:2014kka,chassande2011multimessenger}.

We note that if TianQin is operated within a network of detectors, such as TQ I+II, TQ + LISA, and TQ I+II + LISA, both the expected detection numbers and the precision of source parameter estimation can be significantly improved.
This is true not only for TianQin but also for any other individual detector involved, such as LISA.

Compared to existing literature for similar space-based \ac{GW} missions like LISA/eLISA \cite{Sesana:2016ljz, Sesana:2017vsj, Nishizawa:2016jji, Tamanini:2019usx}), our estimation of the detection number is smaller, but this is due less to the true difference between the detection capabilities of the detectors than to the mass models used in the study.
In particular, we note larger high mass limits in the mass models have been used in earlier works, and this can significantly boost the expected detection numbers because space-borne \ac{GW} detectors are more sensitive to heavier \acp{SBBH}.
By adopting the same setup, the expected detection numbers can be as high as reported in previous studies with LISA.

In summary, TianQin can detect \ac{SBBH} inspirals with good certainty and can measure the corresponding source parameters with impressive precisions.
The analysis from TianQin data alone, as well as from multimessenger observation and multiband \ac{GW} observation, promises great scientific return on astrophysics and fundamental physics related to \acp{SBBH}.

\begin{acknowledgments}
  This work has been supported by the Natural Science Foundation of China (Grants No. 11703098, No. 11805286, No. 91636111, and No. 11690022) and Guangdong Major Project of Basic and Applied Basic Research (Grant No. 2019B030302001). The authors want to thank the anonymous referee for helping us greatly improve the science of this manuscript.
The authors also thank Hai-Tian Wang, Will Farr, Shun-Jia Huang, Peng-Cheng Li, Martin Hendry, and Youjun Lu for helpful discussions.
\end{acknowledgments}

\begin{appendices}
\section*{appendix a: mass distribution model}\label{sec:app1}
\begin{itemize}
\item [(i)] Model {\bf flat-in-log}

The distribution of the masses of both \ac{SBBH} components are independently flat on the logarithmic scale, \\
\begin{equation}
  p(m_{1}, m_{2})\propto\frac{1}{m_{1}m_{2}},
\end{equation}
where $p(m_{1}, m_{2})$ is the probability of \acp{SBBH} with component masses $m_1$ and $m_2$.

\item [(ii)] Model {\bf power law}

The primary mass $m_{1}$ follows a power law distribution while the secondary mass $m_{2}$ follows a uniform distribution,
\begin{align}
  p(m_{1}, m_{2})&\propto \frac{m_{1}^{-\alpha}}{m_{1}-5\msun}\,,\quad \alpha=2.3\,.
\end{align}
\end{itemize}

In these two models, the component masses are bounded by $5\msun < m_{2} < m_{1} < 50\msun$.
\footnote{Note that choice of the upper limit follows \cite{LIGOScientific:2018jsj}, which leads to a more conservative detection number for space-based \ac{GW} detectors compared with other studies adopting earlier, more optimistic upper limit.}

\begin{itemize}
\item [(iii)] Model {\bf A}
\begin{align}\label{model_A}
  p(m_{1},m_{2}|\alpha,\beta_{q})\propto C(m_{1})m_{1}^{-\alpha}q^{\beta_{q}},
\end{align}
where $5\msun\leqslant m_{2}\leqslant m_{1}\leqslant41.6_{-4.3}^{+9.6}\msun$, $q=m_2/m_1$ is the mass ratio, $\alpha=0.4_{-1.9}^{+1.4}$ and $\beta_{q}=0$ are the power law index, and $C(m_{1})$ is a correction factor to make marginalized distribution of $m_1$ follow the power law with index of $\alpha$.

\item [(iv)] Model {\bf B}

  Model follows the same form as model {\bf A}, but with $7.8_{-2.5}^{+1.2}\msun\leqslant m_{2}\leqslant m_{1}\leqslant40.8_{-4.4}^{+11.8}\msun\,$ and $\alpha=1.3_{-1.7}^{+1.4}, \beta_{q}=6.9_{-5.7}^{+4.6}\,$.

\item [(v)] Model {\bf C}

  On top of Model {\bf B}, the possible accumulation of \ac{SBH} due to \ac{PPISN} is characterized by a Gaussian component, and a smooth tail is included in the end, both making the model {\bf C} more realistic,
\begin{align}
  p(m_{1}|\theta)&=\left [ (1-\lambda_{m})A(\theta)m_{1}^{-\alpha}\Theta(m_{\rm max}-m_{1})+\lambda_{m}B(\theta)  \exp \left ( -\frac{(m_{1}-\mu_{m}) ^{2}}{2\sigma_{m}^{2}} \right ) \right ] \nonumber \\
&\times S(m_{1}, m_{\rm min}, \delta m) \\
  p(q|m_{1},\theta)&=C(m_{1},\theta)q^{\beta_{q}}S(m_{2},m_{\rm min},\delta m)
\end{align}
Here $\theta=\{\alpha, m_{\rm max}, m_{\rm min}, \beta_{q}, \lambda_{m}, \mu_{m}, \sigma_{m}, \delta m \}$, whereas $\mu_{m}=29.8_{-7.3}^{+5.8}\msun$ and $\sigma_{m}=6.4_{-4.2}^{+3.2}\msun$ describe the mean and standard deviation of the Gaussian component; $\lambda_{m}=0.3^{+0.4}_{-0.2}$ is the fraction of primary black holes belonging to this Gaussian component; $\alpha=7.1_{-4.8}^{+4.4}, \beta_{q} = 4.5^{+6.6}_{-5.2}, m_{\rm min}=6.9_{-2.8}^{+1.7}\msun$. Functions $A, B, C$ are normalized factors, and function $S(m, m_{min}, \delta m)$, with $\delta m$ being the smooth scale, will smooth the low mass cutoff in the distribution \cite{Talbot:2018cva}.
\end{itemize}

\linespread{1.5}
\begin{table}[h]
\centering
\begin{tabular}{|c|c|c|c|c|c|}
\hline
Model&Flat-in-log&Power law&A&B&C\\
\hline
$\mathcal{R}$(Gpc$^{-3}$yr$^{-1}$)&$19.0_{-8.2}^{+13.0}$&$57.0_{-25.0}^{+40.0}$&$64.0_{-33.0}^{+73.5}$&$53.2_{-28.2}^{+55.8}$&$58.3_{-32.2}^{+72.3}$\\
\hline
\end{tabular}
\caption{The table lists the estimates of merger rate of models flat-in-log, power law, A, B, and C from left to right.}\label{tab:MergRate}
\end{table}

\section*{appendix b: Frequency response for a signal}\label{sec:app2}
In the derivation of Eqs. (\ref{eq:8a}) and (\ref{eq:8b}), we consider the time domain signal induced by a gravitational wave strain, which is given by:
\begin{align}\label{eq:h_I(t)}
  h_{1}(t)=\frac{\sqrt{3}}{2}\left[ F^{+}_{1}(t)h_{+}(t-t_{D})+F^{\times}_{1}(t)h_{\times}(t-t_{D}) \right].
\end{align}
Correspondingly, the frequency domain waveform is the Fourier transform,
\begin{align}\label{eq:FT_hI}
  \widetilde{h}_{1}(f)&=\frac{\sqrt{3}}{2}\{\mathcal{F}[h_{+}(t-t_{D})F_{1}^{+}(t)]+\mathcal{F}[h_{\times}(t-t_{D})F_{1}^{\times}(t)]\} \nonumber \\
  &=\frac{3}{2}\{\mathcal{F}[h_{+}(t-t_{D})]*\mathcal{F}[F_{1}^{+}(t)]+\mathcal{F}[h_{\times}(t-t_{D})]*\mathcal{F}[F_{1}^{\times}(t)]\},
\end{align}
where $\mathcal{F}[\dots]$ denotes the Fourier transformation
 and $*$ represents the convolution
\begin{subequations}\label{FT of Fh}
\begin{align}
  \mathcal{F}[h_{+}(t-t_{D})]&=\int_{-\infty}^{+\infty}{\rm d}t \,h_{+}(t-t_{D})e^{-i2\pi ft} \nonumber \\
  &=e^{-i2\pi ft_{D}}\int_{-\infty}^{+\infty}{\rm d}(t-t_{D}) \,h_{+}(t-t_{D})e^{-i2\pi f(t-t_{D})}\nonumber \\
  &=e^{-i2\pi ft_{D}}\widetilde{h}_{+}(f), \\
  \mathcal{F}[h_{\times}(t-t_{D})]&=e^{-i2\pi ft_{D}}\widetilde{h}_{\times}(f), \\
  \mathcal{F}[F_{1}^{+}(t)]&=\frac{1}{2}(1+\cos^{2}\theta_{S})\mathcal{F}(\cos2\phi_{S})\cos2\psi_{S}-\cos\theta_{S}\mathcal{F}(\sin2\phi_{S})\sin2\psi_{S}, \\
  \mathcal{F}[F_{1}^{\times}(t)]&=\frac{1}{2}(1+\cos^{2}\theta_{S})\mathcal{F}(\sin2\phi_{S})\cos2\psi_{S}+\cos\theta_{S}\mathcal{F}(\sin2\phi_{S})\cos2\psi_{S},
\end{align}
\end{subequations}
where,
\begin{subequations}\label{FT of cos and sin}
\begin{align}
  \mathcal{F}(\cos2\phi_{S})&=\mathcal{F}[\cos2(2\pi f_{0}t+\phi_{S0})] \nonumber \\
  &=\int_{-\infty}^{+\infty}{\rm d}t \,\cos(4\pi f_{0}t+2\phi_{S0})e^{-i2\pi ft} \nonumber \\
  &=\frac{1}{2}e^{i2\phi_{S0}}\int_{\infty}^{+\infty}{\rm d}t\, e^{i2\pi(2f_{0}-f)t} +\frac{1}{2}e^{-i2\phi_{S0}}\int_{\infty}^{+\infty}{\rm d}t\, e^{-i2\pi(2f_{0}+f)t} \nonumber \\
  &=\frac{1}{2}e^{i2\phi_{S0}}\delta(f-2f_{0})+\frac{1}{2}e^{-i2\phi_{S0}}\delta(f+2f_{0}), \\
  \mathcal{F}(\sin2\phi_{S})&=\mathcal{F}[\sin2(2\pi f_{0}t+\phi_{S0})] \nonumber \\
  &=-\frac{i}{2}e^{i2\phi_{S0}}\delta(f-2f_{0})+\frac{i}{2}e^{-i2\phi_{S0}}\delta(f+2f_{0}).
\end{align}
\end{subequations}
Substituting Eqs. (\ref{FT of Fh}) and (\ref{FT of cos and sin}) into the $\mathcal{F}[h_{+}(t-t_{D})]*\mathcal{F}[F_{1}^{+}(t)]$ and $\mathcal{F}[h_{\times}(t-t_{D})]*\mathcal{F}[F_{1}^{\times}(t)]$, we obtain 
\begin{subequations}
\begin{align}
\mathcal{F}[h_{+}(t-t_{D})F^{+}_{1}(t)]&=\frac{1}{4}(1+\cos^{2}\theta_{S})\left[e^{2i\zeta_{1}(f-2f_{0})}\widetilde{h}_{+}(f-2f_{0})+e^{-2i\zeta_{2}(f+2f_{0})}\widetilde{h}_{+}(f+2f_{0})\right]\cos2\psi_{S} \nonumber \\
  &-\frac{i}{2}\cos\theta_{S}\left[-e^{2i\zeta_{1}(f-2f_{0})}\widetilde{h}_{+}(f-2f_{0})+e^{-2i\zeta_{2}(f+2f_{0})}\widetilde{h}_{+}(f+2f_{0})\right]\sin2\psi_{S}, \\
  \mathcal{F}[h_{\times}(t-t_{D})F^{\times}_{1}(t)]&=\frac{1}{4}(1+\cos^{2}\theta_{S})\left[e^{2i\zeta_{1}(f-2f_{0})}\widetilde{h}_{\times}(f-2f_{0})+e^{-2i\zeta_{2}(f+2f_{0})}\widetilde{h}_{\times}(f+2f_{0})\right]\sin2\psi_{S} \nonumber\\
  &+\frac{i}{2}\cos\theta_{S}\left[-e^{2i\zeta_{1}(f-2f_{0})}\widetilde{h}_{\times}(f-2f_{0})+e^{-2i\zeta_{2}(f+2f_{0})}\widetilde{h}_{\times}(f+2f_{0})\right]\cos2\psi_{S},
\end{align}
\end{subequations}
\end{appendices}
where $\zeta_{1}(f)=\phi_{S0}-\pi ft_{D}$, $\zeta_{2}(f)=\phi_{S0}+\pi ft_{D}$.


\bibliography{reference}
\end{document}